%% file: distrwelfaretest.tex
\documentclass[11pt, english]{article}
\usepackage[T1]{fontenc}
\usepackage[utf8]{inputenc}
\usepackage{amsmath} 
\usepackage{amsthm} 
\usepackage{amssymb}
\usepackage{mathrsfs}
\usepackage{esint}
\usepackage{babel}
\usepackage{bbm}   
\usepackage{graphicx}
\usepackage{color}
\usepackage{subfig}
\usepackage{multirow}
\usepackage{dsfont}
\usepackage{lscape}
\usepackage{ctable}

\usepackage{pdfpages}

\usepackage[margin=3cm]{geometry}

\usepackage{latexsym} % For a weak convergence symbol
\usepackage{enumitem} % For special enumerated lists.
\usepackage[round]{natbib}

\usepackage{setspace}
\newtheorem{theorem}{Theorem}[section]
\newtheorem{prop}[theorem]{Proposition} 
\newtheorem{lemma}[theorem]{Lemma}
\newtheorem{corollary}[theorem]{Corollary}

\theoremstyle{definition}
\newtheorem{deff}[theorem]{Definition} 
\newtheorem{remark}[theorem]{Remark} 
\newtheorem{example}[theorem]{Example}

	\global\long\def\sbr#1{\left[#1\right] }
	
	\global\long\def\rbr#1{\left(#1\right)}
	
	\global\long\def\R{\mathbb{R}}

	\global\long\def\dd#1{\textnormal{d}#1}

  \global\long\def\calF{\mathcal{F}}
  \global\long\def\calG{\mathcal{G}}

  \global\long\def\calM{\mathcal{M}}
  \global\long\def\calP{\mathcal{P}}
  \global\long\def\calU{\mathcal{U}}
  \global\long\def\calX{\mathcal{X}}
  
  \global\long\def\one{\mathbf{1}}
  \global\long\def\bbD{\mathbb{D}}
  \global\long\def\bbE{\mathbb{E}}
  \global\long\def\bbF{\mathbb{F}}
  \global\long\def\bbG{\mathbb{G}}
  \global\long\def\bbL{\mathbb{L}}
  \global\long\def\bbU{\mathbb{U}}
  
  \global\long\def\cw{\leadsto}
  
  \global\long\def\ns{\varnothing}
  \global\long\def\prob#1{P \left\{ #1 \right\}}
  \global\long\def\ex#1{\mathrm{E} \left[ #1 \right]}

  \newcommand{\argmax}{\operatornamewithlimits{argmax}}

\title{Loss aversion and the welfare ranking of policy interventions\thanks{
The authors are grateful to Pedro Carneiro, Dan Hamermesh, Hide Ichimura, Rados\l{}aw Kurek, Essie Maasoumi, Piotr Mi\l{}o\'{s}, Magne Mogstad, Jim Powell, Jo\~{a}o Santos Silva, Tiemen Woutersen, and seminar participants at the University of Chicago, University of Wisconsin-Madison, USC, MSU, University of Montreal, University of California Berkeley, University of Arizona, 28th annual meeting of the Midwest Econometrics Group, 35th Meeting of the Canadian Econometric Study Group, and the 3rd edition of the Rio-Sao Paulo Econometrics Workshop %, and 2017 Latin American Meeting of the Econometric Society 
for useful comments and discussions regarding this paper. Martyna Kobus acknowledges the support of the National Science Centre, Poland, the grant no. 2020/39/B/HS4/03131. Marta Schoch provided excellent research assistance. Computer programs to replicate the numerical analyses are available from the authors. This work was made possible by the facilities of the Shared Hierarchical Academic Research Computing Network (SHARCNET:www.sharcnet.ca) and Compute/Calcul Canada.  All the remaining errors are ours.}}
\author{Sergio Firpo\footnote{Insper, Sao Paulo, Brazil. E-mail: \texttt{\
firpo@insper.edu.br}} \and 
Antonio F. Galvao\footnote{Department of Economics, Michigan State University, East Lansing, USA. E-mail: \texttt{\ agalvao@msu.edu}} \and Martyna Kobus\footnote{Institute of Economics, Polish Academy of Sciences, Warsaw, Poland. E-mail:  \texttt{\ mkobus@inepan.waw.pl}} \and Thomas Parker\footnote{Department of Economics, University of Waterloo, Waterloo, Canada. E-mail: \texttt{\ tmparker@uwaterloo.ca}} \and Pedro Rosa-Dias\footnote{Department of Economics and Public Policy, Imperial College Business School, Imperial College London, UK. E-mail: \texttt{\ p.rosa-dias@imperial.ac.uk}}
}

\begin{document} 
\thispagestyle{empty}
\maketitle

\vspace{-1.25cm}

\begin{abstract}
\begin{spacing}{1}
This paper develops theoretical criteria and econometric methods to rank policy interventions in terms of welfare when individuals are loss-averse. Our new criterion for ``loss aversion-sensitive dominance'' defines a weak partial ordering of the distributions of policy-induced gains and losses. It applies to the class of welfare functions which model individual preferences with non-decreasing and loss-averse attitudes towards changes in outcomes. We also develop new semiparametric statistical methods to test loss aversion-sensitive dominance in practice, using nonparametric plug-in estimates; these allow inference to be conducted through a special resampling procedure. Since point-identification of the distribution of policy-induced gains and losses may require  strong assumptions, we extend our comparison criteria, test statistics, and resampling procedures to the partially-identified case. We illustrate our methods with a simple empirical application to the welfare comparison of alternative income support programs in the US.
  
\end{spacing}
\end{abstract} 

\vspace{-0.50cm}



\newpage

\begin{quote}
\scriptsize
\em We suffer more, ... when we fall from a better to a worse situation, than we ever enjoy when we rise from a worse to a better.
\end{quote}
\begin{flushright}
\scriptsize Adam Smith, The Theory of Moral Sentiments
\end{flushright}

%\begin{flushright}
%\scriptsize\em
%
%``We suffer more, ..., when we fall from a better to a worse situation, than we ever enjoy when we rise from a worse to a better.'' 
%
%\em  Adam Smith, The Theory of Moral Sentiments
%\end{flushright}

\section{Introduction}
\setcounter{page}{1}

Policy interventions often generate heterogeneous effects, giving rise to gains and losses to different individuals and sectors of society. Classically, the welfare ranking of policy interventions, conducted under the Rawlsian principle of the ``veil of ignorance'', has deemed such gains and losses irrelevant: all policies that produce the same marginal  distribution of outcomes should be considered equivalent for the purpose of welfare analysis.  \citep{Atkinson70,Roemer98, Sen00}.
However, more recent approaches have focused precisely on how different individuals are affected by a given policy  \citep{HeckmanSmith98, CarneiroHansenHeckman01}. Our paper relates to this latter approach and focuses on an important characteristic of the individual valuation of policy-induced effects: loss aversion.\footnote{Loss aversion is a well established empirical regularity, documented in a wide variety of contexts \citep{KahnemanTversky79, SamuelsonZeckhauser88, TverskyKahneman91, RabinThaler:01, Rick11}} 

There are two main reasons why loss aversion can be important for the welfare ranking of policy interventions.  First, as shown in \citet{CarneiroHansenHeckman01}, the individual gains and losses caused by a policy have important political economy consequences.  Public support for that policy, and for the authorities that implement it, depends on the balance of gains and losses experienced and valued by different individuals in the electorate. In this context, there is mounting empirical evidence indicating that the electorate often exhibits loss aversion. This aversion to losses among constituents, in turn, drives the actions of policy makers, as documented in situations as diverse as government support to the steel industry in US trade policy, and President Trump's attempted repeal of the Affordable Care Act \citep{FreundOezden08, AlesinaPassarelli19}. 
 
Second, political economy aside, there are important situations where policy makers have strong normative reasons for incorporating loss-aversion in their own welfare ranking of public policies. This has been proposed in a range of fields. In a recent example, \citet{Eyal20} shows that the Hippocratic principle of “first, do no harm” has led US and EU policy makers to delay the development of Covid-19 vaccines by rejecting human challenge trials, that involve purposefully infecting a small number of vaccine trial volunteers. In this situation, policy makers placed more weight on the potential harm to a small number of volunteers, than on the vast potential benefits of accelerating the availability of a vaccine to large swathes of the population. Similarly, in the context of minimum wage legislation, \citet{Mankiw14} proposes that policy makers should be loss-averse in their approach to policy evaluation and adopt a ``first, do no harm'' principle: ``\emph{As I see it, the minimum wage and the Affordable Care Act are cases in point. Noble as they are in aspiration, they fail the do-no harm test. An increase in the minimum wage would disrupt some deals that workers and employers have made voluntarily}''. Along the same lines, in the context of development economics, scholars such as \citet{Easterly09} have proposed a ``first, do no harm'' approach to foreign aid interventions in developing countries.   
 
In this paper, we extend the toolkit available for the evaluation of policy interventions in contexts where it is sensible to incorporate loss-aversion in the welfare ranking of policy interventions. We are not proposing that \textit{all policies} should be ranked under loss aversion sensitive criteria. Instead, we provide a methodology for conducting such a ranking in cases where aversion to losses is likely to be important. As discussed above, this may be the case either because policy makers are genuinely loss-averse, or because they know that the individuals exposed to the policy are so, and this leads them to incorporate loss aversion in policy ranking. To this end, our paper develops new testable criteria and econometric methods to rank distributions of individual policy effects, from a welfare standpoint, incorporating loss aversion. We make two main contributions to the literature. 

Our first contribution is to propose loss aversion-sensitive criteria for the welfare ranking of policies. We adopt the standard welfare function approach \citep{Atkinson70}: alternative policies are compared based on a welfare ranking, where social welfare is an additively separable and symmetric function of individuals' outcomes. It is well established that, for non-decreasing utility functions, this is equivalent to first-order stochastic dominance (FOSD) over distributions of policy outcomes. Analogously, our ranking is based on social \emph{value} functions, which are additively separable and symmetric functions of individual \emph{gains and losses}.  We show that the social value function ranking with non-decreasing and loss-averse value functions \citep{TverskyKahneman91} is equivalent to a new concept we call loss aversion-sensitive dominance (LASD) over distributions of policy-induced gains and losses. FOSD requires that the cumulative distribution function of the dominated distribution lies everywhere above the cumulative distribution of the dominant distribution. In contrast, under LASD, the dominated cumulative distribution function must lie sufficiently above the dominant distribution function for \emph{losses} such that the probability of potential losses cannot be compensated by a higher probability for potential gains. This is a consequence of loss-aversion. Except for the special case of a \emph{status quo} policy (i.e. a policy of no change) where FOSD and LASD coincide, generally, as we show, LASD can be used to compare policies that are indistinguishable for FOSD.\footnote{The literature on stochastic dominance is vast and spans economics and mathematics - we refer the reader to, e.g., \citet{ShakedShanthikumar94} and \citet{Levy16} for a review. When dominance curves cross, higher order or inverse stochastic dominance criteria have been proposed. The former involves conditions on higher (typically third and fourth) order derivatives of utility function (e.g. \citet{Fishburn80}, \citet{Chew83} to which \citet{EeckhoudtSchlesinger06} provided interesting interpretation, whereas the latter is related to the rank-dependent theory originally proposed by \citet{Weymark81} and \citet{Yaari87, Yaari88}, where social welfare functions are weighted averages of ordered outcomes with weights decreasing with the rank of the outcome (see \citet{AabergeHavnesMogstad18} for a recent refinement of this theory). }

The LASD criterion relies on gains and losses, which under standard identification conditions can be considered \emph{treatment effects}. It is well known that the point identification of the distribution of treatment effects may require implausible theoretical restrictions such as rank invariance of potential outcomes \citep{HeckmanSmithClements97}. We thus extend our LASD criteria to a partially-identified setting and establish a sufficient condition to rank alternative policies under partial identification of the distributions of their effects. We use Makarov bounds \citep{Makarov82, Rueschendorf82, FrankNelsenSchweizer87} to bound the distribution of treatment effects when the joint pre and post-policy outcome distribution is unknown. This provides a testable criterion that can be used in practice, since the marginal distribution functions from samples observed under various treatments can usually be identified and Makarov bounds only rely on marginal information for their identification.

Our second contribution is to develop statistical inference procedures to practically test the loss averse-sensitive dominance condition using sample data. We develop statistical tests for both point-identified and partially-identified distributions of outcomes. The test procedures are designed to assess, uniformly over the two outcome distributions, whether one treatment dominates another in terms of the LASD criterion.  Specifically, we suggest Kolmogorov-Smirnov and Cram\'er-von Mises test statistics that are applied to nonparametric plug-in estimates of the LASD criterion mentioned above. Inference for these statistics uses specially tailored resampling procedures.  We show that our procedures control the size of tests for all probability distributions that satisfy the null hypothesis.  Our tests are related to the literature on inference for stochastic dominance represented by, e.g., \citet{LintonSongWhang10, LintonMaasoumiWhang05, BarrettDonald03} and references cited therein. \citet{LintonMaasoumiWhang05} is an important contribution because in addition to developing tests for stochastic dominance of arbitrary order, they propose a Prospect Theory stochastic dominance test. Their test is intended for inferring dominance among a different family of value functions than ours, namely, the focus is on risk loving for gains and risk aversion for losses (i.e. so called S-shapedness, \citet{KahnemanTversky79}), but not on loss aversion. %DavidsonDuclos13, Maasoumi01, DavidsonDuclos00, DardanoniForcina99, Anderson96, KaurPrakasaRaoSingh94, Klecanetal91, McFadden89}. 
We contribute to the literature by developing tests for loss averse-sensitive dominance, which are an alternative to standard stochastic dominance tests.  Our tests widen the variety of comparisons available to empirical researchers to other criteria that encode important qualitative features of agent preferences.

%Our contribution also fits into the broader literature on inference for many moment inequalities~--- see for example \citet{ChernChetKato18, LeeSongWhang18, AndrewsShi17}.

%Our econometric approach extends existing stochastic dominance testing procedures. 
The LASD criterion results in a functional inequality that depends on marginal distribution functions, and we adapt existing techniques from the literature on testing functional inequalities to test for LASD. However, in comparison with stochastic dominance tests, verifying LASD with sample data presents technical challenges for both the point- and partially-identified cases. The criterion that implies LASD of one distribution over another is more complex than the standard FOSD criterion, and hence requires a significant extension of existing procedures to justify the use of inference about loss averse-sensitive dominance with a nonparametric plug-in estimator of the LASD criterion.

In particular, the problem with using existing stochastic dominance techniques is that the mapping of distribution functions to a testable criterion is nonlinear and ill-behaved. A dominance test inherently requires uniform comparisons be made, and tractable analysis of its distribution demands regularity, in the form of differentiability, of the map between the space of distribution functions and the space of criterion functions. However, the map from pairs of distribution functions to the LASD criterion function is not differentiable.  Despite this complication, we show that supremum- or $L_2$-norm statistics applied to this function are just regular enough that, with some care, resampling can be used to conduct inference.

For practical implementation, we propose an inference procedure that combines standard resampling with an estimate of the way that test statistics depend on underlying data distributions, building on recent results from \citet{FangSantos19}.  We contribute to the literature on directionally differentiable test statistics with a new test for LASD. Recent contributions to this literature include, among others, \citet{HongLi18, ChetverikovSantosShaikh18, ChoWhite18, ChristensenConnault19, FangSantos19, CattaneoJanssonNagasawa17} and \citet{MastenPoirier17}.  

When distributions are only partially identified by bounds, the situation is more challenging. The current state of the literature on Makarov bounds focuses on pointwise inference for bound functions (see, e.g., \citet{FanPark10,FanPark12}, \citet{FanGuerreZhu17}, and \citet{FirpoRidder08,FirpoRidder19}).  However, the LASD criterion requires a uniform comparison of bound functions, and the map from distribution functions to Makarov bound functions is also not smooth.  Fortunately the problem has a similar solution to the point-identified LASD test. The resulting $L_2$- and supremum-norm statistics allow us to conduct inference for LASD in the partially identified case using functions that bound the relevant CDFs.  The details are included in the online Supplemental Appendix.

%Finally, this paper also relates to the strand of literature that develops methods to estimate the optimal treatment assignment policy that maximizes a social welfare function. Recent developments can be found in \citet{Manski04}, \citet{Dehejia05}, \citet{HiranoPorter09}, \citet{Stoye09}, \citet{BhattacharyaDupas12}, \citet{Tetenov12}, \citet{KitagawaTetenov18, KitagawaTetenov19}, among others. These papers focus on the decision-theoretic properties and procedures that map empirical data into treatment choices.  In this context, our paper is most closely related to \citet{Kasy16}, which focuses on welfare rankings of policies rather than optimal policy choice.  

%\citet{Kasy16} analyzed treatment choice for a class of social welfare functions including rank-dependent social welfare, and developed a set of identification results that map distributions of observed data into partial orderings (identified rankings) of policies. 
%In addition, \citet{Kasy16} considered a linear approximation of the rank-dependent welfare function around a status-quo policy to discuss (partial) identification of a welfare-improving local policy change. 
%Our approach differs from \citet{Kasy16} in several aspects: (a) we consider loss aversion as an axiom on decision maker preference; (b) we do not specify any functional form for the social value function; (c) the statistical analysis does not rely on any parametric specification.

We illustrate the practical use of our proposed criteria and tests with a very simple empirical application using data from \citet{BitlerGelbachHoynes06}. This aims at exemplifying the use of our approach, rather than developing a fully fledged empirical investigation.  We show that, in the case of a policy with gainers and losers, the use of our loss aversion-sensitive evaluation criteria may lead to a ranking of policy interventions that differs from that obtained when their outcomes are compared using stochastic dominance.

The rest of the paper is organized as follows. Section \ref{sec:basic} presents the basic definitions and notation and defines loss aversion-sensitive dominance. Section \ref{sec:ptsd} develops testable criteria for loss aversion-sensitive dominance. Section \ref{sec:inference} proposes statistical inference methods for LASD using sample observations. Section \ref{sec:applic} illustrates our methodology using a very simple empirical application that uses data from the experimental evaluation of a well-known welfare policy reform in the US. Section \ref{sec:conclusion} concludes. Our first appendix includes auxiliary results and definitions; our second one collects proof of the results in the paper.

\section{Loss aversion-sensitive dominance} \label{sec:basic}

In this section, we propose a novel dominance relation for ordering policies under the assumption that social decision makers consider the distribution of individual gains and losses under different policy scenarios. We call this criterion Loss Aversion-Sensitive Dominance (LASD). 

Consider a random variable $X$ with cumulative distribution function $F$. Let $\mathscr{F}$ be the set of cumulative distribution functions with bounded support $\calX$.  We maintain the assumption throughout that $F \in \mathscr{F}$.  The bounded support assumption is made to avoid technical conditions on tails of distribution functions. The aim of this paper is to provide theoretical criteria and econometric methods to rank policy interventions under LASD. The decision maker's goal is to compare policies $A$ and $B$ using the distribution functions of $X_A$ and $X_B$, labeled $F_A$ and $F_B$. 
Because the random variables $X_A$ and $X_B$ represent gains and losses due to the enactment of policies $A$ and $B$, they represent a change between agents' pre-treatment and post-treatment outcomes.  
To this end, let $Z_0, Z_A$ and $Z_B$ represent an agent's potential outcome under the \textit{status quo}, treatment $A$ or treatment $B$, so that we may write $X_A = Z_A - Z_0$ and $X_B = Z_B - Z_0$.  We assume that the variables $(Z_0, Z_A, Z_B)$ have marginal distribution functions $(G_0, G_A, G_B)$.

It may be assumed that $(F_A, F_B)$ are identified, or that only $(G_0, G_A, G_B)$ are identified.  The former case is related to a point-identified model, and the latter to a partially-identified one. Theoretical results for both situations are shown here.  Inference for the first situation is considered in this paper, while partially-identified inference results under the second situation are considered in the online Supplemental Appendix.

For example, in the empirical illustration considered in Section~\ref{sec:applic}, we observe quarterly household income before the enactment of a new welfare program, which we assume to represent realizations of $Z_0$.  Next, we observe income for some households under a continuation of the old program, identifying those as realizations of $Z_A$, and income for other households under a new welfare program, labeling those observations as $Z_B$ realizations.  If policymakers are interested in how one household's earnings evolve over time either by staying with the old program or switching to the new program, then the levels of $Z_A$ and $Z_B$ are not of primary interest, rather the changes represented by $X_A$ and $X_B$ are (and given the longitudinal nature of our data, it is natural to assume that $X_A$ and $X_B$ are identified).  Our tests, detailed in Section~\ref{sec:inference}, use the null hypothesis that a household exhibiting loss-aversion would prefer to switch to the new welfare program, and search for evidence to the contrary.

%\textcolor{red}{(ORIGINAL) Suppose a random variable $X$ describes individual gains and losses, and $X$ has cumulative distribution function $F$, and let $\mathscr{F}$ be the set of cumulative distribution functions with bounded support $\calX$.  We maintain the assumption throughout that $F \in \mathscr{F}$.  The bounded support assumption is made to avoid technical conditions on tails of distribution functions. The aim of this paper is to provide theoretical criteria and econometric methods to rank policy interventions under LASD. Before proceeding we define some distributions that will be used for formalizing the methods and make comparisons between policies $A$ and $B$. The agents' current outcomes are represented by the random variable $Z_0$ which has marginal distribution function $G_0$.  Two other random variables, $Z_A$ and $Z_B$, describe outcomes under policies $A$ and $B$.  Assume their marginal distribution functions are $G_A$ and $G_B$ respectively. The gains and losses due to policies $A$ and $B$ are defined by the random variables $X_A = Z_A - Z_0$ and $X_B = Z_B - Z_0$. The decision maker's goal is to compare policies $A$ and $B$ using the distribution functions of $X_A$ and $X_B$, labeled $F_A$ and $F_B$.}

The decision maker has preferences over $X$ (not $Z$) that are represented via a continuous function.
\begin{deff}[Social Value Function (SVF)] \label{def:svf}
  Suppose random variable $X$ has CDF $F \in \mathscr{F}$ and let $W: \mathscr{F} \rightarrow \R$ denote the following \emph{social value function} 
\begin{equation} \label{eq:svf}
  W(F) = \int_\calX v(x) \dd F(x), 
\end{equation} 
where $v: \calX \rightarrow \R$ is called a \emph{value function}.\footnote{Formally speaking we have $W_v(F)$ but we suppress the subscript $v$ for expositional brevity.} 
\end{deff}

The social value function defined above is the value assigned to the distribution of $X$ by a social planner that uses the value function $v$ to convert gains and losses into a measure of well-being \citep{GajdosWeymark12}. This value function $v$ does not have to coincide with any individual's $v$  in the population: as mentioned in the Introduction, the social planner is averse to individual losses either because individuals are loss-averse themselves (political economy motivation), or because she holds normative views that imply her loss aversion towards $X$. In either case, $v$ will exhibit loss-aversion, i.e. there is asymmetry in the valuation of gains and losses, where losses are weighed more heavily than gains of equal magnitude. Furthermore, $v$ assigns negative value to losses and positive value to gains and is non-decreasing. These properties are formally listed in the next definition.\footnote{This standard interpretation of the social welfare function can be further extended. For example, individuals may be uncertain about their counterfactual outcome and form an expectation of $v(\cdot)$ given $z_0$. Then we write
$W(F) = \int \int v(x, z_0) \dd F_{X|Z_0} (x|z_0) \dd F_{Z_0}(z_0)$.  Denoting a new value function $v^*(z_0) = \int v(x, z_0) \dd F_{X|Z_0} (x|z_0)$, i.e. an expected value for a given $z_0$, $W(F)$ is as in Definition \ref{def:svf}.}$^{,}$\footnote{An interesting direction for future research is to axiomatize the class of social value functions and possibly develop measures of loss aversion based on this class. Some inspiration for axiomatization may come from the inequality and poverty measurement literature. For example, ratio scale invariance (i.e. proportional changes to the units in which gains and losses are measured do not matter), may be a powerful axiom in obtaining a specific functional form.}

\begin{deff}[Properties of the value function] \label{deff:properties}
  The value function $v: \calX \rightarrow \R$ is differentiable and satisfies:
\begin{enumerate} 
  \item Disutility of losses and utility of gains: $v(x) \leq 0$ for all $x < 0$, $v(0) = 0$ and $v(x) \geq 0$ for all $x > 0$.
  \item Non-decreasing: $v'(x) \geq 0$ for all $x$.
  \item Loss-averse: $v'(-x) \geq v'(x)$ for all $x>0$.
\end{enumerate} 
\end{deff}

The properties in Definition \ref{deff:properties} are typically assumed in Prospect Theory together with the additional requirement of S-shapedness of value function, which we do not consider (see, e.g., p. 279 of \citet{KahnemanTversky79}). Assumptions 1 and 2 are standard monotone increasing conditions. Assumption 3 expresses the idea that ``losses loom larger than corresponding gains'' and is a widely accepted definition of loss aversion \citep[p.303]{TverskyKahneman92}. It is a stronger condition than the one considered by \citet{KahnemanTversky79}. 

The following form of $W(F)$ will be useful in subsequent definitions and results. 
\begin{prop} \label{prop:svf_alt}
  Suppose that $F \in \mathscr{F}$ and $v$ is differentiable.  Then
\begin{equation} \label{svf2}
W(F) = -\int\displaylimits_{x \in \calX: x \leq 0} v'(x)F(x)\dd x + \int\displaylimits_{x \in \calX: x > 0} v'(x)(1-F(x))\dd x.
\end{equation}
\end{prop}

Assume that the decision maker's social value function $W$ depends on $v$ which satisfies Definition~\ref{deff:properties}, and she wishes to compare random variables $X_{A}$ and $X_{B}$ which represent gains and losses under two policies labeled $A$ and $B$. The decision maker prefers $X_A$ over $X_B$ if she evaluates $F_A$ as better than $F_B$ using her SVF~--- specifically, $X_A$ is preferred to $X_B$ if and only if $W(F_A) \geq W(F_B)$, where $W$ is defined in Definition~\ref{def:svf}. Please note that $X_A$ is preferred to $X_B$ for \emph{every} $v$ that is described by Definition~\ref{deff:properties}. This is what makes dominance conditions robust criteria for comparing distributions. This idea is formalized below. 

\begin{deff}[Loss Aversion-Sensitive Dominance]\label{def:PT-FSD}
  Let $X_A$ and $X_B$ have distribution functions respectively labeled $F_A, F_B \in \mathscr{F}$.  If $W(F_A) \geq W(F_B)$ for all value functions $v$ that satisfy Definition~\ref{deff:properties}, we say that $F_A$ dominates $F_B$ in terms of \emph{Loss Aversion-Sensitive Dominance}, or LASD for short, and we write $F_A \succeq_{LASD} F_B$.
\end{deff}

In the next section we relate this theoretical definition to a more concrete condition that depends on the cumulative distribution functions of the outcome distributions, $F_A$ and $F_B$.

\section{Testable criteria for loss aversion-sensitive dominance}\label{sec:ptsd}
In this section we formulate testable conditions for evaluating distributions of gains and losses in practice. We propose criteria that indicate whether one distribution of gains and losses dominates another in the sense described in Definition~\ref{def:PT-FSD}.

Recall that $Z_0, Z_A$ and $Z_B$ represent an outcome before or after a policy takes effect, while $X_A$ and $X_B$ represent a change from a pre-policy state to an outcome under a policy.  The challenge of comparing variables $X_A$ and $X_B$ is well known in the treatment effects literature: because $X_A$ and $X_B$ are defined by differences between the $Z_k$, $F_A$ and $F_B$ depend on the joint distribution of $(Z_0, Z_A, Z_B)$, which may not be observable without restrictions imposed by an economic model.  In subsection \ref{sec:point_iden} we abstract from specific identification conditions and discusses LASD under the assumption that $F_A$ and $F_B$ are identified.  In subsection \ref{sec:partial_iden} we work with a partially identified case where only the marginal distribution functions $G_0$, $G_A$ and $G_B$ are identified and no restrictions are made to identify $F_A$ and $F_B$.

\subsection{The case of point-identified distributions}\label{sec:point_iden}
The LASD concept in Definition \ref{def:PT-FSD} requires that one distribution is preferred to another over a class of social value functions and is difficult to test directly.  The following result relates the LASD concept to a criterion which depends only on marginal distribution functions and orders $F_A$ and $F_B$ according to the class of SVFs allowed in Definition~\ref{deff:properties}.
%The class of SVF in Definition \ref{def:PT-FSD} is unobservable, but the following result relates it to a criterion for the distributions which is observable and thus implements the ordering of $F$ according to $W$.
In this section we assume that $F_A, F_B \in \mathscr{F}$ are point identified.  This may result from a variety of econometric restrictions that deliver identification and are the subject of a large literature.

\begin{theorem} \label{thm:ptfsd}
  Suppose that $F_A, F_B \in \mathscr{F}$.  The following are equivalent:
  \begin{enumerate}
    \item $F_A \succeq_{LASD} F_B$.
    \item For all $x \geq 0$, $F_A$ and $F_B$ satisfy
      \begin{equation} \label{eq:ptfsd}
        F_{B}(-x) - F_{A}(-x) \geq \max\{ 0, F_{A}(x)-F_{B}(x)\}.
      \end{equation} 
    \item For all $x \geq 0$, $F_A$ and $F_B$ simultaneously satisfy
      \begin{equation} \label{eq:ptfsd1}
        F_{A}(-x)-F_{B}(-x)\leq 0
      \end{equation} 
      and 
      \begin{equation} \label{eq:ptfsd2}     
        \rbr{1-F_{A}(x)}-F_{A}(-x) \geq \rbr{1-F_{B}(x)}-F_{B}(-x).
      \end{equation} 
  \end{enumerate}
\end{theorem}

Theorem~\ref{thm:ptfsd} provides two different conditions that can be used to verify whether one distribution of gains and losses dominates the other in the LASD sense.\footnote{LASD is a partial order. Over losses, \eqref{eq:ptfsd1} is a partial order because FOSD is a partial order. For the tail condition \eqref{eq:ptfsd2} checking transitivity we have $\rbr{1-F_{A}(x)}-F_{A}(-x) \geq \rbr{1-F_{B}(x)}-F_{B}(-x), \rbr{1-F_{B}(x)}-F_{B}(-x) \geq \rbr{1-F_{C}(x)}-F_{C}(-x)$, and $\rbr{1-F_{A}(x)}-F_{A}(-x) \geq \rbr{1-F_{C}(x)}-F_{C}(-x)$. If $F_{A}(-x)-F_{B}(-x)=0$ then $F_{A}(-x)=F_{B}(-x)$ and using it in \eqref{eq:ptfsd2} gives anti-symmetry.} These criteria compare the outcome distributions by examining how the distribution functions $(F_A, F_B)$ assign probabilities to gains and losses of all possible magnitudes. The particular way that they make a comparison is related to the relative importance of gains and losses.  Consider condition~\eqref{eq:ptfsd}.  For the distribution of $X_B$ to be dominated, its distribution function must lie above the distribution of $X_A$ \emph{for losses}. 
$X_B$ can be dominated by $X_A$ in the LASD sense even when gains under $X_A$ do not dominate $X_B$ for gains~--- that is, when $F_{A}(x)-F_{B}(x) \geq 0$ for some $x \geq 0$~--- as long as this lack of dominance in gains is compensated by  sufficient dominance of $X_A$ over $X_B$ in the losses region. This is a consequence of the asymmetric treatment of gains and losses.  Conditions~\eqref{eq:ptfsd1} and~\eqref{eq:ptfsd2} jointly express the same idea, but they help to understand how gains and losses are treated asymmetrically in condition~\eqref{eq:ptfsd}. In the losses region, condition~\eqref{eq:ptfsd1} is a standard FOSD condition. This is a consequence of loss aversion; note that in the extreme case where only losses matter, we would have \eqref{eq:ptfsd1}. In the gains region, dominance has to be sufficiently large so that under $X_A$, the probability of gains minus the probability of losses (of magnitude $x$ or larger) is no smaller than the corresponding difference for $X_B$.\footnote{We leave 1s on both sides of inequality~\eqref{eq:ptfsd2} for this interpretation to be more evident.}   Inequality~\eqref{eq:ptfsd} combines the two inequalities represented by~\eqref{eq:ptfsd1} and~\eqref{eq:ptfsd2} into a single equation.

%The result of Theorem~\ref{thm:ptfsd} is related to FOSD comparisons.  Recall that using the standard FOSD criterion, for $X_B$ to be dominated, the  distribution function $F_B$ lies everywhere above $F_A$, the cumulative distribution of the dominant distribution.

It is interesting to note that LASD has one property in common with FOSD, namely, a higher mean is a necessary condition for both types of dominance. This follows directly from Definitions \ref{def:svf} and \ref{deff:properties} by using $v(x) = x$.
\begin{lemma} \label{corr:means} 
  If $F_{A} \succeq_{LASD} F_{B}$ then $\ex{X_A} \geq \ex{X_B}$.  
\end{lemma}

Note that FOSD cannot rank two distributions that have the same mean~--- that is, if $F_A \succeq_{FOSD} F_B$ and $\ex{X_A} = \ex{X_B}$, then $F_A = F_B$.  This is not the case for LASD, as the next example demonstrates.  Therefore, for example, equation~\eqref{eq:ptfsd} may still be used to differentiate between two distributions with the same average effect.

\begin{example} \label{ex:counterexamp_PTFSD}
  Consider the family of uniform distributions on $[-1-y, -y]\cup [y, y+1]$ indexed by $y > 0$ and denote the corresponding member distribution functions $F_y$. The family of such distributions have mean zero and $F_{y} \succeq_{LASD} F_{y'}$ whenever $y < y'$. Indeed, note that
  \begin{equation*}
    W(F_y) = \frac{1}{2}\rbr{\int_{-1-y}^{-y} v(z) \dd z + \int_{y}^{1+y} v(z) \dd z}
  \end{equation*}
  and thus for any $v$ which is loss-averse (see Definition~\ref{deff:properties}) we have
\begin{align*}
  \frac{\dd}{\dd y} W(F_y) &= \frac{1}{2}\rbr{v(-1-y) -v(-y) + v(1+y) - v(y)} \\
  {} &= -\int_{-1-y}^{-y} v'(z) \dd z + \int_{y}^{1+y} v'(z) \dd z \\
	&=  \int_{y}^{1+y} \rbr{v'(z) - v'(-z)} \dd z\leq 0.	
\end{align*}
\end{example}

It is important to note that LASD is a concept that is specialized to the comparison of distributions that represent gains and losses.  Standard FOSD is typically applied to the distribution of outcomes in levels without regard to whether the outcomes resulted from gains or losses of agents relative to a pre-policy state~--- in our notation, $G_A$ and $G_B$ are typically compared with FOSD, instead of $F_A$ and $F_B$. FOSD applied to post-policy levels may or may not coincide with LASD applied to changes. This means that even when a strong condition such as FOSD holds for final outcomes, if one took into account how agents value gains and losses it may turn out that the dominant distribution is no longer a preferred outcome.  One could apply the FOSD rule to compare distributions of income changes, which implies LASD applied to changes, because FOSD applies to a broader class of value functions.  However, this type of comparison would ignore agents' loss aversion, the important qualitative feature that LASD accounts for.  The following example shows that the analysis of outcomes in levels using FOSD need not correspond to any LASD ordering of outcomes in changes.

\begin{example}
Let $Z_0$ represent outcomes before policies $A$ or $B$.  Suppose $Z_0$ is distributed uniformly over $\{0, 1, 2, 3\}$. Policy $A$ assigns post-policy outcomes depending on the realized $Z_0$ according to the schedule 
$$
Z_A = \begin{cases}
3 & \text{if } \{Z_0 = 0\}\\
2 & \text{if } \{Z_0 = 1\}\\
0 & \text{if } \{Z_0 = 2\}\\
1 & \text{if } \{Z_0 = 3\}.
\end{cases}
$$
Therefore the distribution of $X_A = Z_A - Z_0$ is $P\{X_A = -2\} = 1/2$, $P\{X_A = 1\} = P\{X_A = 3\} = 1/4$.  Meanwhile, policy $B$ maintains the status quo: $X_B = Z_B - Z_0 = 0$ with probability 1.

Distributions $G_A$ and $G_B$ are the same, thus FOSD holds between $Z_A$ and $Z_B$. However, there is no loss aversion-sensitive dominance between $X_A$ and $X_B$. Indeed, we can find two value functions that fulfill the conditions of Definition \ref{def:PT-FSD} but order $X_A$ and $X_B$ differently. For example, take $v_1(x) = x^3$. Then $3 = \int v_1(x) dF_A(x) > \int v_1(x) dF_B(x) = 0$. Next let $v_2(x) = \text{sgn}(x)|x|^{1/3}$. Then $-0.02 \approx \int v_2(x) dF_A(x) < \int v_2(x) dF_B(x) \approx 0$. 
\end{example}

%The LASD concept is different from using FOSD to compare a distribution of gains and losses to a single reference income level \citet{Levy16}. In such a case, the relationship between cumulative distributions (i.e. first-order stochastic dominance) on gains and losses is preserved when a reference income is added to a gain/loss. Here, however, the reference distribution may also exhibit gains and losses.
%So far we argued that FSD and LASD are different concepts. 

In the previous example, policy $B$ left pre-treatment outcomes unchanged, or in other words, maintained a \textit{status quo} condition~--- we had $X_B = Z_B - Z_0 \equiv 0$.  Suppose generally that $X_B$ has a distribution that is degenerate at $0$.  Then $F_B(x) = 0$ for all $x < 0$ and $F_B(x) = 1$ for all $x \geq 0$. We define this as a \textit{status quo} policy distribution, labelled $F_{SQ}$. When comparison is between a distribution $F_A$ and $F_{SQ}$, LASD and standard FOSD are equivalent. The distribution that dominates $F_{SQ}$ is necessarily only gains.

\begin{corollary}\label{corr:statusquo}
  Suppose that $F_A \in \mathscr{F}$ and $F_B = F_{SQ}$.  Then $F_{A} \succeq_{LASD} F_{SQ} \iff F_{A} \succeq_{FOSD} F_{SQ}$.
\end{corollary} 

\subsection{The case of partially-identified distributions}\label{sec:partial_iden}

In many situations of interest the cumulative distribution functions of gains and losses, $F_A$ and $F_B$, are not point identified without a model of the relationship between $X_A$ and $X_B$. However, the marginal distributions of outcomes in levels under different policies, represented by the variables $Z_0$, $Z_A$ and $Z_B$, may be identified. Without information on the dependence between potential outcomes, we can still make some more circumscribed statements with regard to dominance based on bounds for the distribution functions.
This section studies the LASD dominance criterion to the case that distribution functions $F_A$ and $F_B$ are only partially identified.

A number of authors have considered functions that bound the distribution functions $F_A$ and $F_B$.  Taking $X_A$ as an example, \emph{Makarov bounds} \citep{Makarov82, Rueschendorf82, FrankNelsenSchweizer87} are two functions $L$ and $U$ that satisfy $L(x) \leq F_A(x) \leq U(x)$ for all $x \in \R$, depend only on the marginal distribution functions $G_0$ and $G_A$ and are pointwise sharp~--- for any fixed $x$ there exist some $Z_0^*$ and $Z_A^*$ such that the resulting $X_A^* = Z^*_A - Z^*_0$ has a distribution function at $x$ that is equal one of $L(x)$ or $U(x)$.  \citet{WilliamsonDowns90} provide convenient definitions for these bound functions.  For any two distribution functions $G_1, G_2$, define 
\begin{align*}
  L(x, G_1, G_2) &= \sup_{u \in \R} \left( G_2(u) - G_1(u - x) \right) \\
  U(x, G_1, G_2) &= \inf_{u \in \R} \left( 1 + G_2(u) - G_1(u - x) \right).
\end{align*}
For convenience define the policy-specific bound functions for $F_k$, $k \in \{A, B\}$ and all $x \in \R$, which depend on the marginal CDFs $G_0$ and $G_k$, by
\begin{align}
  L_k(x) &= L(x, G_0, G_k) \label{lowerk} \\
  {} &= \sup_{u \in \calX} (G_k(u) - G_0(u - x)) \notag \\
  U_k(x) &= U(x, G_0, G_k). \label{upperk} \\
  {} &= 1 + \inf_{u \in \calX} (G_k(u) - G_0(u - x)) \notag
\end{align}
Using these definitions we obtain a sufficient and a necessary condition for LASD when only bound functions of the treatment effects distribution functions are observable. The next theorem formalizes the result.
\begin{theorem} \label{thm:boundsfsd}
  Suppose that $G_0, G_A, G_B \in \mathscr{F}$ and define the bound functions using formulas~(\ref{lowerk}) and~(\ref{upperk}) for $k \in \{A, B\}$.
  \begin{enumerate}
    \item If for all $x \geq 0$,
      \begin{equation} \label{ptfsdsuff}
        L_{B}(-x)-U_{A}(-x)\geq \max\{0,U_{A}(x) - L_{B}(x)\},
      \end{equation}
      then $F_A \succeq_{LASD} F_B$ holds.
    \item If $F_A \succeq_{LASD} F_B$  holds, then for all $x \geq 0$
      \begin{equation} \label{ptfsdnec}
        U_{B}(-x)-L_{A}(-x) \geq \max\{0, L_{A}(x)-U_{B}(x)\}.
      \end{equation}	
  \end{enumerate}
\end{theorem}

Theorem \ref{thm:boundsfsd} is an analog of Theorem \ref{thm:ptfsd} and shows what effect the loss of point identification has on the relationship between dominance and conditions on the CDFs. In particular, one loses a simple ``if and only if'' characterization that depends on CDFs. Instead, LASD implies a necessary condition using some bound functions, while a different sufficient condition using other bound functions implies LASD.  Inference using the necessary condition shown in Theorem \ref{thm:boundsfsd} is discussed in the online Supplemental Appendix. We remark that there may exist other features of the joint data distribution that do not depend only on pointwise features of the CDFs of changes and would result in a necessary and sufficient condition for LASD under partial identification. That is an interesting open question but is beyond the scope of this paper.

When the comparison is with the \textit{status quo} distribution, the partially identified conditions simplify.  Corollary~\ref{corr:boundsfsd} below shows what can be learned about LASD from bound functions in the partially identified case.

%Theorem \ref{thm:boundsfsd} is an implication of Theorem \ref{thm:ptfsd} in the partially-identified setting. Both Theorems \ref{thm:ptfsd} and \ref{thm:boundsfsd} will play important parts in the inference procedures discussed in the next Section.

%When the comparison is with the \textit{status quo} distribution, the partially identified conditions simplify.  Corollary~\ref{corr:boundsfsd} below is an extension of Corollary~\ref{corr:statusquo} to the partially identified case.

\begin{corollary}\label{corr:boundsfsd}
  Suppose that $F_B = F_{SQ}$ and that $G_0, G_A \in \mathscr{F}$. Define the bound functions $U_A$ and $L_A$ using formulas~(\ref{lowerk}) and~(\ref{upperk}).  Then $U_A(-x) = 0$ for all $x \geq 0 \Rightarrow F_{A} \succeq_{LASD} F_{SQ}$ and $F_{A} \succeq_{LASD} F_{SQ} \Rightarrow L_A(-x) = 0$ for all $x \geq 0$.
\end{corollary}

\section{Inferring loss aversion-sensitive dominance} \label{sec:inference}

In this section we propose statistical inference methods for the loss aversion-sensitive dominance (LASD) criterion discussed in previous sections. We consider the null and alternative hypotheses
\begin{equation} \label{hypotheses_general}
  \begin{aligned}
    H_0 &: F_A \succeq_{LASD} F_B \\
    H_1 &: F_A \not\succeq_{LASD} F_B.
  \end{aligned}
\end{equation}
Under the null hypothesis \eqref{hypotheses_general} policy $A$ dominates $B$ in the LASD sense, similar to much of the literature on stochastic dominance.  It is a simplification of the hypotheses considered for several potential policies discussed in \citet{LintonMaasoumiWhang05}, who test whether one policy is maximal, and the techniques developed below could be extended to compare several policies in the same way in a straightforward manner.\footnote{\citet{LintonMaasoumiWhang05} consider a test for Prospect Theory by  testing whether the integral of one CDF dominates the other. This paper considers a different approach in which we impose loss aversion on the value function, and then derive testable conditions on the CDFs.}  The null hypothesis above represents the assumption that policy $A$ is preferred by agents in the LASD sense. Rejection of the null implies that there is significant evidence for ambiguity in the ordering of the policies by LASD. Unfortunately, a drawback of the proposed procedure is that rejection of the null does not inform one about which sort of value function $v$ results in a rejection.  Strong orderings of policies can result in more information, although they constrain $v$ by construction, and such exploration is left for future research.\footnote{There is also another strand of literature that develops methods to estimate the optimal treatment assignment policy that maximizes a social welfare function. Recent developments can be found in \citet{Manski04}, \citet{Dehejia05}, \citet{HiranoPorter09}, \citet{Stoye09}, \citet{BhattacharyaDupas12}, \citet{Tetenov12}, \citet{KitagawaTetenov18, KitagawaTetenov19}, among others. These papers focus on the decision-theoretic properties and procedures that map empirical data into treatment choices.  In this literature, our paper is most closely related to \citet{Kasy16}, which focuses on welfare rankings of policies rather than optimal policy choice.}

%We use the dominance criteria discussed in Theorems~\ref{thm:ptfsd} and \ref{thm:boundsfsd} to design nonparametric tests for $H_0$.  Because the LASD hypothesis is translated into functional inequalities, which we discuss below, tests must be conducted uniformly over all $x \geq 0$.  This uniformity in $x$ and features of the LASD conditions present a challenge for inference.

We consider tests for this null hypothesis given sample data observed under two different identification assumptions.  We start with the case where one can directly observe samples $\{X_{Ai}\}_{i=1}^{n_A}$ and $\{X_{Bi}\}_{i=1}^{n_B}$ which represent agents' gains and losses, or in other words, we simply assume that the distribution functions of $X_A$ and $X_B$ are point-identified and their distribution functions can be estimated using the empirical distribution functions from two samples. Next we extend these results to the partially-identified case where no assumption about the joint distribution of potential outcomes under either treatment is made.  In this case, we assume that three samples are observable, $\{Z_{0i}\}_{i=1}^{n_0}$, $\{Z_{Ai}\}_{i=1}^{n_A}$ and $\{Z_{Bi}\}_{i=1}^{n_B}$, representing outcomes under a control or pre-policy state and outcomes under policies $A$ and $B$. Then tests are based on plug-in estimates for bounds for $X_A = Z_A - Z_0$ and $X_B = Z_B - Z_0$.

We consider distribution functions as members of the space of bounded functions on the support $\calX \subseteq \R$, denoted $\ell^\infty(\calX)$, equipped with the supremum norm, defined for $g: \R^k \rightarrow \R^\ell$ by $\| g \|_\infty = \max_j \{ \sup_{x \in \R^k} |g_j(x)| \}$.  For real numbers $x$ let $(x)^+ = \max\{0, x\}$.  Given a sequence of bounded functions $\{g_n\}_n$ and limiting random element $g$ we write $g_n \cw g$ to denote weak convergence in $(\ell^\infty, \| \cdot \|_\infty)$ in the sense of Hoffman-J\o rgensen \citep{vanderVaartWellner96}.

\subsection{Inferring dominance from point identified treatment distributions} \label{sec:pointID}

In this subsection we suppose that the pair of marginal distribution functions $F = (F_A, F_B)$ is identified. In the Online Supplemental Appendix C, we provide results extending the dominance tests to the case that distribution functions $F_A$ and $F_B$ are only partially identified. 

\subsubsection{Test statistics}
To implement a test of the hypotheses \eqref{hypotheses_general} we employ the results of Theorem~\ref{thm:ptfsd} to construct maps of $F$ into criterion functions that are used to detect deviations from the hypothesis $H_0$.  Specifically, recalling that $(x)^+ = \max\{0, x\}$, for the point-identified case we examine maps $T_1: (\ell^\infty(\R))^2 \rightarrow \ell^\infty(\R_+)$ and $T_2: (\ell^\infty(\R))^2 \rightarrow (\ell^\infty(\R_+))^2$, defined for each $x \geq 0$ by 
\begin{equation} \label{t1def}
  T_1(F)(x) = (F_A(x) - F_B(x))^+ + F_A(-x) - F_B(-x)
\end{equation}
and
\begin{equation} \label{t2def}
  T_2(F)(x) = \begin{bmatrix} F_A(-x) - F_B(-x) \\ F_A(x) - F_B(x) + F_A(-x) - F_B(-x) \end{bmatrix}.
\end{equation}
Functions $T_1(F)$ and $T_2(F)$ are designed so that large positive values will indicate a violation of the null.  Taking $T_1$ as an example, Theorem~\ref{thm:ptfsd} states that $W(F_A) \geq W(F_B)$ if and only if $F_B(-x) - F_A(-x) \geq (F_A(x) - F_B(x))^+$ for all $x \geq 0$, so tests can be constructed by looking for $x$ where $T_1(F)(x)$ becomes significantly positive.  We will refer to $T_j$ as maps from pairs of distribution functions to another function space, and also refer to them as functions.

The hypotheses~(\ref{hypotheses_general}) can be rewritten in two equivalent forms, depending on whether one uses $T_1$ or $T_2$ to transform distribution functions: letting $\calX \subseteq \R_+$ be an evaluation set, we have
\begin{equation}\label{hypotheses_T1}
  \begin{aligned}
  H_0^{(1)} &: T_1(F)(x) \leq 0, \quad \text{for all } x \in \calX, \\
  H_1^{(1)} &: T_1(F)(x) > 0, \quad \text{for some } x \in \calX
  \end{aligned}
\end{equation}
and
\begin{equation}\label{hypotheses_T2}
  \begin{aligned}
  H_0^{(2)} &: T_2(F)(x) \leq 0_2, \quad \text{for all } x \in \calX, \\
  H_1^{(2)} &: T_2(F)(x) \not\leq 0_2, \quad \text{for some } x \in \calX.
  \end{aligned}
\end{equation}
In the second set of hypotheses $0_{2}$ is a two-dimensional vector of zeros and inequalities are taken coordinate-wise.

The next step in testing the hypotheses \eqref{hypotheses_T1} and \eqref{hypotheses_T2} is to estimate $T_1(F)$ and $T_2(F)$.  Let $\bbF_n = (\bbF_{An}, \bbF_{Bn})$ denote the pair of marginal empirical distribution functions, that is, $\bbF_{kn}(x) = \frac{1}{n_k} \sum_{i=1}^{n_k} \one\{X_{ki} \leq x\}$ for $k \in \{A, B\}$.  These are well-behaved estimators of the components of $F$. Letting $n = n_A + n_B$, standard empirical process theory shows that $\sqrt{n} (\bbF_n - F)$ converges weakly to a Gaussian process under weak assumptions \citep[Example 19.6]{vanderVaart98}.  In order to conduct inference for loss aversion-sensitive dominance, we use plug-in estimators $T_j(\bbF_n)$ for $j \in \{1, 2\}$.  See Remark~\ref{rem:compute} in Appendix~\ref{app:hadamard} for details on the computation of these functions.

In order to detect when $T_j(\bbF_n)$ is significantly positive, we consider statistics based on a one-sided supremum norm or a one-sided $L_2$ norm over $\calX$.  Kolmogorov-Smirnov (i.e., supremum norm) type statistics are
\begin{align}
  V_{1n} &= \sqrt{n} \sup_{x \in \calX} (T_1(\bbF_n)(x))^+ \label{stats_first} \\
  V_{2n} &= \sqrt{n} \max \left\{ \sup_{x \in \calX} (T_{21}(\bbF_n)(x))^+, \sup_{x \in \calX} (T_{22}(\bbF_n)(x))^+ \right\}.
\end{align}
Meanwhile Cram\'er-von Mises (or $L_2$ norm) test statistics are defined by
\begin{align}
  W_{1n} &= \sqrt{n} \left( \int_\calX \left( (T_1(\bbF_n)(x))^+ \right)^2 \dd x \right)^{1/2}, \\
  W_{2n} &= \sqrt{n} \left( \int_\calX \left( (T_{21}(\bbF_n)(x))^+ \right)^2 + \left( (T_{22}(\bbF_n)(x))^+ \right)^2 \dd x \right)^{1/2}. \label{stats_last}
\end{align}
Because all the CDFs used in these statistics belong to $\mathscr{F}$, distributions with bounded support, the integrands in the $L_2$ statistics are square-integrable.

\subsubsection{Limiting distributions}
We wish to establish the limiting distributions of $V_{jn}$ and $W_{jn}$, for $j\in\{1,2\}$, under the null hypothesis $H_0: F_A \succeq_{LASD} F_B$.  Two challenges arise when considering these test statistics.  First, the form of the null hypothesis as a functional inequality to be tested uniformly over $\calX$ is a source of irregularity. Let the joint probability distribution of $(X_A, X_B)$ be denoted by $P$. Because the null hypothesis, $F_A \succeq_{LASD} F_B$, is a functional weak inequality the asymptotic distributions of the test statistics $V_j$ and $W_j$  may depend on features of $P$.  This is referred to as \emph{non-uniformity in $P$} in \citep{LintonSongWhang10, AndrewsShi13}, and requires attention when resampling.

Second, due to the pointwise maximum function in its definition, $T_1$ is too irregular as a map from the data to the space of bounded functions to establish a limiting distribution for the empirical process $\sqrt{n}(T_1(\bbF_n) - T_1(F))$ using conventional statistical techniques.  In contrast, $T_2$ is a linear map of $F$, which implies that $\sqrt{n}(T_2(\bbF_n) - T_2(F))$ has a well-behaved limiting distribution in $(\ell^\infty(\R_+))^2$.\footnote{The issues of a general lack of differentiability of functions arrived at by marginal optimization and a solution for inference based on directly characterizing the behavior of test statistics applied to such functions are studied in more generality in~\citet{FirpoGalvaoParker21}. However, we highlight that the tests described here are extensions of the results of that paper and are specifically tailored to this application.}
%the tests described here do not fit the general results of that paper and are specifically tailored to this application.}

Despite the above challenges, we show that $V_{jn}$ and $W_{jn}$ (for $j \in \{1, 2\}$) have well-behaved asymptotic distributions, and furthermore, that the limiting random variables satisfy $V_1 \sim V_2$ and $W_1 \sim W_2$.  This is an important result because it is the foundation for applying bootstrap techniques for inference.  Before stating the formal assumptions and asymptotic properties of the tests, we discuss the two difficulties mentioned above in more detail.

The limiting distributions of $V_{jn}$ and $W_{jn}$ statistics depend on features of $P$.
Let $\calP_0$ be the set of distributions $P$ such that $F_A \succeq_{LASD} F_B$.  These are distributions with marginal distribution functions $F$ such that $T_j(F)(x) \leq 0$ for all $x \geq 0$.  To discuss the relationship between these sets of distributions and test statistics, we relabel the two coordinates of the $T_2$ function as 
\begin{equation}
  m_1(x) = F_A(-x) - F_B(-x) 
\end{equation}
and
\begin{equation}
  m_2(x) = F_A(-x) - F_B(-x) + F_A(x) - F_B(x). 
\end{equation}
When $P \in \calP_0$, both $m_1(x) \leq 0$ and $m_2(x) \leq 0$ for all $x \geq 0$.  

More detail is required about the behavior of the two coordinate functions to determine the limiting distributions of $V_{jn}$ and $W_{jn}$ statistics.  For $L_2$-norm statistics $W_{1n}$ and $W_{2n}$, we define the following relevant subdomains of $\calX$, which collect the arguments in the interior of $\calX$ where $m_1$ or $m_2$ are equal to zero:
\begin{align}
  \calX_0^1(P) &= \{x \in \text{int} \calX : m_1(x) = 0\} \\
  \calX_0^2(P) &= \{x \in \text{int} \calX : m_2(x) = 0\}.
\end{align}
Denote $\calX_0(P) \subseteq \calX$ as the set of $x$ where $T_1(F)(x) = 0$ or at least one coordinate of $T_2(F)$ equals $0$ for probability distribution $P$.  As will be seen below, $\calX_0(P)$ is the same for both the $T_1$ and $T_2$ functions, and when it is non-empty, test statistics have a nondegenerate distribution.  Following \citet{LintonSongWhang10}, we call $\calX_0(P)$ the contact set for the distribution $P$.  Given the above definitions, under the null hypothesis we can write
\begin{equation*}
  \calX_0(P) = \calX_0^1(P) \cup \calX_0^2(P).
\end{equation*}
On the other hand, the supremum-norm statistics $V_{1n}$ and $V_{2n}$ need a different family of sets, namely the sets of $\epsilon$-maximizers of $m_1$ and $m_2$.  For any $\epsilon \geq 0$ and $k \in \{1, 2\}$, let
\begin{equation}
	\calM^k(\epsilon) = \left\{ x \in \calX : m_k(x) \geq \sup_{x \in \calX} m_k(x) - \epsilon \right\}.
\end{equation}

An important subset of $\calP_0$ are those $P$ for which test statistics have nontrivial limiting distributions under the null hypothesis~--- that is, not degenerate at 0, which occurs when there is some $x$ such that $T_j(F)(x) = 0$ (note that there are no $x$ such that $T_j(F)(x) > 0$ when $P \in \calP_0$).  Define $\calP_{00} \subset \calP_0$ to be the set of all $P$ such that $\calX_0(P) \neq \ns$.  If $P \in \calP_0 \backslash \calP_{00}$ then $\calX_0(P) = \ns$ and because the distribution satisfies the null hypothesis, $F_A$ strictly dominates $F_B$ everywhere and the criterion functions $T_j$ are strictly negative over $\calX$.  When $P \in \calP_0 \backslash \calP_{00}$, test statistics have asymptotic distributions that are degenerate at zero because test statistics will detect that policy $A$ is strictly better that $B$ over all of $\calX$.  When $P \in \calP_{00}$, $T_j(F)$ is zero over $\calX_0(P)$ and test statistics have a nontrivial asymptotic distribution over $\calX_0(P)$.  Thus, when $F_A \succeq_{LASD} F_B$, the asymptotic behavior of test statistics depends on whether $P \in \calP_{00}$ or $P \in \calP_0 \backslash \calP_{00}$.  Note that when $P \in \calP_{00}$, we have $\lim_{\epsilon \searrow 0} \calM^k(\epsilon) = \calX_0^k(P)$ (that is, nonstochastic convergence in the sense of Painlev\'e-Kuratowski, see, e.g., \citet[p. 111]{RockafellarWets98}) for whichever coordinate function actually achieves the maximal value zero.

The second challenge for testing is related to the scaled difference $\sqrt{n}(T_1(\bbF_n) - T_1(F))$ as $n$ grows large.  Hadamard differentiability is an analytic tool used to establish the asymptotic distribution of nonlinear maps of the empirical process.  Definition~\ref{def:diff} in Appendix~\ref{app:hadamard} provides a precise statement of the concept.  When a map is Hadamard differentiable~--- for example $T_2$, which is linear as a map from $(\ell^\infty(\R))^2$ to $(\ell^\infty(\R_+))^2$ and is thus trivially differentiable~--- the functional delta method can be applied to describe its asymptotic behavior as a transformed empirical process, and a chain rule makes the analysis of compositions of several Hadamard-differentiable maps tractable.  Also, the Hadamard differentiability of a map implies resampling is consistent when this map is applied to the resampled empirical process \citep[Theorem 23.9]{vanderVaart98}~--- so, for example, the distribution of resampled criterion processes $\sqrt{n}(T_2(\bbF_n^*) - T_2(\bbF_n))$ is a consistent estimate of the asymptotic distribution of $\sqrt{n}(T_2(\bbF_n) - T_2(F))$ in the space $\ell^\infty(\R_+)$.  On the other hand, consider the $T_1$ map.  The pointwise Hadamard directional derivative of $T_1(f)(x)$ at a given $x \geq 0$ in direction $h(x) = (h_A(x), h_B(x))$ is
\begin{equation} \label{t1prime}
  T_{1f}'(h)(x) = \begin{cases} h_A(x) - h_B(x) + h_A(-x) - h_B(-x), & f_A(x) > f_B(x) \\ (h_A(x) - h_B(x))^+ + h_A(-x) - h_B(-x), & f_A(x) = f_B(x) \\ h_A(-x) - h_B(-x), & f_A(x) < f_B(x) \end{cases}.
\end{equation}
This map, thought of as a map between function spaces, $(\ell^\infty(\R))^2$ and $\ell^\infty(\R_+)$, is not differentiable because the scaled differences $(T_1(f)(x) - T_1(f + th_t)(x)) / t$ converge to the above derivative at each point $x$, but may not converge uniformly in $\R_+$.  Despite the lack of differentiability of the map $F \mapsto T_1(F)$, we show in Lemma~\ref{lem:t1diff} in Appendix~\ref{app:hadamard} that the maps $F \mapsto V_1$ and $F \mapsto W_1$ are Hadamard directionally differentiable, which implies these maps are just regular enough that existing statistical methods can be applied to their analysis.  Later in this section we apply the resampling technique recently developed in \citet{FangSantos19} along with this directional differentiability to describe hypothesis tests using $V_{1n}$ or $W_{1n}$.

Having discussed the difficulties in the relationship between distributions and test statistics, we turn to assumptions on the observations.  In order to conduct inference using either $T_1(\bbF_n)$ or $T_2(\bbF_n)$ we make the following assumptions.
\begin{enumerate}[label=\textbf{A\arabic*}]
  \item \label{assumptionA_first} The observations $\{X_{Ai}\}_{i=1}^{n_A}$ and $\{X_{Bi}\}_{i=1}^{n_B}$ are iid samples and independent of each other and are continuously distributed with marginal distribution functions $F_A$ and $F_B$ respectively.
  \item \label{assumptionA_last} Let the sample sizes $n_A$ and $n_B$ increase in such a way that $n_k / (n_A + n_B) \rightarrow \lambda_k$ as $n_A, n_B \rightarrow \infty$, where $0 < \lambda_k < 1$ for $k \in \{A, B\}$.  Define $n = n_A + n_B$.
  %\item The evaluation set $\calX \subseteq \mathrm{supp}(F_A) \cup \mathrm{supp}(F_B)$, the union of the supports of the marginal distribution functions $F_A$ and $F_B$.
\end{enumerate}
%The assumptions used above could be generalized in several directions.  \citet{Klecanetal91} discuss generalizations on the independence assumptions made in the first assumption above (stationarity and $\alpha$-mixing within samples, exchangeability between samples); these can be made if need be without changing the main features of our results.

Under these assumptions we establish the asymptotic properties of the test statistics under the null and fixed alternatives.  Under the above assumptions, there is a Gaussian process $\calG_F$ such that $\sqrt{n}(\bbF_n - F) \cw \calG_F$.  We denote each coordinate process $\calG_{F_A}$ and $\calG_{F_B}$, and for convenience define two transformed processes: for each $x \geq 0$ let
\begin{align}
	\calG_1(x) &= \calG_{F_A}(-x) - \calG_{F_B}(-x) \label{calg1_def} \\
	\calG_2(x) &= \calG_{F_A}(x) - \calG_{F_B}(x) - \calG_{F_A}(-x) + \calG_{F_B}(-x). \label{calg2_def}
\end{align}
These will be used in the theorem below.

\begin{theorem} \label{thm:obs_teststats}
  Make assumptions~\ref{assumptionA_first}-\ref{assumptionA_last}.  Define the limiting Gaussian processes $\calG_1$ and $\calG_2$ as above.  Then:
  \begin{enumerate}
    \item Suppose that $P \in \calP_{00}$.  As $n \rightarrow \infty$, $V_{1n} \cw V_1$ and $W_{1n} \cw W_1$, where
	\begin{equation*}
	V_1 \sim \max \left\{ 0, \sup_{x \in \calX_0^1(P)} \calG_1(x) \cdot \one \left\{ \sup_{x \in \calX} m_1(x) = 0 \right\}, \sup_{x \in \calX_0^2(P)} \calG_2(x) \cdot \one \left\{ \sup_{x \in \calX} m_2(x) = 0 \right\} \right\}
	\end{equation*}
      and
      \begin{equation*}
        W_1 \sim \left( \int_{\calX_0^1(P)} \left( \left( \calG_1(x) \right)^+ \right)^2 \dd x + \int_{\calX_0^2(P)} \left( \left( \calG_2(x) \right)^+ \right)^2 \dd x \right)^{1/2}.
      \end{equation*}
    \item Suppose that $P \in \calP_{00}$.  As $n \rightarrow \infty$, $V_{2n} \cw V_2$ and $W_{2n} \cw W_2$, where $V_2 \sim V_1$ and $W_2 \sim W_1$.
    \item Suppose that $P \in \calP_0 \backslash \calP_{00}$ for $j = 1$ or $2$.  As $n \rightarrow \infty$, $\prob{V_{jn} > \epsilon} \rightarrow 0$ and $\prob{W_{jn} > \epsilon} \rightarrow 0$ for all $\epsilon > 0$.
    \item Suppose that $P \not\in \calP_0$.  As $n \rightarrow \infty$, $\prob{V_{jn} > c} \rightarrow 1$ and $\prob{W_{jn} > c} \rightarrow 1$ for all $c \geq 0$ for $j = 1$ or $2$.
  \end{enumerate}
\end{theorem}

Theorem~\ref{thm:obs_teststats} derives the asymptotic properties of the proposed test statistics. Parts 1 and 2 establish the weak limits of $V_{jn}$ and $W_{jn}$ for $j\in\{1,2\}$ when the null hypothesis is true.  Recall that when $P \in \calP_{00}$, $\lim_{\epsilon \searrow 0} \calM^k(\epsilon) = \calX_0^k(P)$, which is why $\calM^k(\epsilon)$ terms are absent in the first part of the theorem.  Remarkably, the test statistics using $T_1$ and $T_2$ criterion processes have the same asymptotic behavior despite the different appearances of the underlying processes and the irregularity of $T_1$. Part 3 shows that the statistics are asymptotically degenerate at zero when the contact set is empty, that is, when $P$ lies on the interior of the null region. Part 4 shows that the test statistics diverge when data comes from any distribution that does not satisfy the null hypothesis.  

The limiting distributions described in Part 1 of Theorem~\ref{thm:obs_teststats} are not standard because the distributions of the test statistics depend on features of $P$ through the $\calX_0(P)$ terms in each expression. Therefore, to make practical inference feasible, we suggest the use of resampling techniques below.

\subsubsection{Resampling procedures for inference}
The proposed test statistics have complex limiting distributions. In this subsection, we present resampling procedures to estimate the limiting distributions of both $V_{jn}$ and $W_{jn}$ for $j\in\{1,2\}$ under the assumption that $P \in \calP_{00}$.  Naive use of bootstrap data generating processes in the place of the original empirical process suffers from distortions due to discontinuities in the directional derivatives of the maps that define the distributions of the test statistics.  In finite samples the plug-in estimate will not find, for example, the region where $F_A(x) - F_B(x) = 0$, where the derivatives exhibit discontinuous behavior.  Our procedure involves making estimates of the derivatives involved in the limiting distribution and a standard exchangeable bootstrap routine, as proposed in \citet{FangSantos19}.\footnote{Given a set of weights $\{W_i\}_{i=1}^n$ that sum to one and are independent of $\{X_i\}_{i=1}^n$, the exchangeable bootstrap measure is a randomly-weighted measure that puts mass $W_i$ at observed sample point $X_i$ for each $i$.  This encompasses, for example, the standard bootstrap, $m$-of-$n$ bootstrap and wild bootstrap.  See Section 3.6.2 of \citet{vanderVaartWellner96} for more specific details.}

In order to estimate contact sets, define a sequence of constants $\{a_n\}$ such that $a_n \searrow 0$ and $\sqrt{n}a_n \rightarrow \infty$ and let $\hat{m}_{1n}(x) = \bbF_{An}(-x) - \bbF_{Bn}(-x)$ and $\hat{m}_{2n}(x) = \bbF_{An}(-x) - \bbF_{Bn}(-x) + \bbF_{An}(x) - \bbF_{Bn}(x)$.  Then for $W_j$ statistics define estimated contact sets by
\begin{align}
  \hat{\calX}_0^1 &= \{x \in \text{int} \calX : |\hat{m}_{1n}(x)| \leq a_n\} \label{tcontact_est} \\
  \hat{\calX}_0^2 &= \{x \in \text{int} \calX : |\hat{m}_{2n}(x)| \leq a_n\}.
\end{align}
When both sets are empty, replace both estimates by $\calX$, as suggested in \citet{LintonSongWhang10} to ensure nondegenerate bootstrap reference distributions.  Meanwhile, for $V_j$ statistics define estimated $\epsilon$-maximizer sets.  For a sequence of constants $\{b_n\}$ such that $b_n \searrow 0$ and $\sqrt{n} b_n \rightarrow \infty$, let
\begin{align}
	\hat{\calM}^1 &= \{x \in \calX : \hat{m}_{1n}(x) \geq \max \hat{m}_{1n}(x) - b_n \}, \\
	\hat{\calM}^2 &= \{x \in \calX : \hat{m}_{2n}(x) \geq \max \hat{m}_{2n}(x) - b_n \}.
\end{align}

Although the null hypothesis may imply that the maximum $m_1(x)$ is zero, the above formulas use the maximum of the sample analog without setting its maximum equal to zero, which is important for ensuring non-empty set estimates.  Using these estimates, the distributions of $V_1$ and $W_1$ can be estimated from sample data (recall that Part~2 of Theorem~\ref{thm:obs_teststats} asserts that these are the same distributions as those of $V_2$ and $W_2$).  We conducted simulation experiments to choose these parameters using a few simulated data-generating processes, which are briefly discussed in the appendix in the context of simulations that suggest that the resulting tests have correct size and good power.  Scaling the estimated processes by their pointwise standard deviation functions when estimating contact sets as in~\citet{LeeSongWhang18} might result in better performance when distribution functions are evaluated near their tails, but we leave that rather complex topic for future research.

\noindent \textbf{Resampling routine to estimate the distributions of $V_{jn}$ and $W_{jn}$ for $j = 1, 2$:}
\begin{enumerate}
  \item \label{tcontact} If using a Cram\'er-von Mises statistic, given a sequence of constants $\{a_n\}$, estimate the contact sets $\hat{\calX}_0^1$ and $\hat{\calX}_0^2$.  If using a Kolmogorov-Smirnov statistic, given a sequence of constants $\{b_n\}$, estimate the $b_n$-maximizer sets of $\hat{m}_{1n}$ and $\hat{m}_{2n}$.
\end{enumerate}
Next repeat the following two steps for $r = 1, \ldots, R$:
\begin{enumerate}
    \setcounter{enumi}{1}
  \item Construct the resampled processes
  \begin{align*}
  	\calF_{r1n}^*(x) &= \sqrt{n} \Big( \bbF^*_{An}(-x) - \bbF^*_{Bn}(-x) - \bbF_{An}(-x) + \bbF_{Bn}(-x) \Big) \\
	\calF_{r2n}^*(x) &= \sqrt{n} \Big( \bbF^*_{An}(-x) - \bbF^*_{Bn}(-x) - \bbF_{An}(-x) + \bbF_{Bn}(-x) \\
	&+ \bbF^*_{An}(x) - \bbF^*_{Bn}(x) - \bbF_{An}(x) + \bbF_{Bn}(x) \Big)
  \end{align*}
  using an exchangeable bootstrap.
  \item Calculate the resampled test statistic.  Letting $\hat{k} = \argmax_k \{\sup_{x \geq 0} \hat{m}_{kn}(x)\}$ and $\{c_n\} \searrow 0$ satisfy $\sqrt{n}c_n \rightarrow \infty$, calculate
\begin{equation} \label{vstar_def}
	V_{rn}^* = \begin{cases} 
	\left( \max_{x \in \hat{\calM}^{\hat{k}}} \calF_{r\hat{k}n}^*(x) \right)^+ & | \max \hat{m}_{1n} - \max \hat{m}_{2n} | > c_n \\
	\max \left\{ 0, \max_{x \in \hat{\calM}^1} \calF_{r1n}^*(x), \max_{x \in \hat{\calM}^2} \calF_{r2n}^*(x) \right\} & | \max \hat{m}_{1n} - \max \hat{m}_{2n} | \leq c_n
	\end{cases}
\end{equation}
or
\begin{equation} \label{wstar_def}
	W_{rn}^* = \left( \int_{\hat{\calX}_0^1} \left( \left( \calF_{r1n}^*(x) \right)^+ \right)^2 \dd x + \int_{\hat{\calX}_0^2} \left( \left( \calF_{r2n}^*(x) \right)^+ \right)^2 \dd x \right)^{1/2}.
\end{equation}
\end{enumerate}
Finally,
\begin{enumerate}
    \setcounter{enumi}{3}
  \item Let $\hat{q}_{V^*}(1-\alpha)$ and $\hat{q}_{W^*}(1-\alpha)$ be the $(1-\alpha)^{\text{th}}$ sample quantile from the bootstrap distributions of $\{V_{rn}^*\}_{r=1}^R$ or $\{W_{rn}^*\}_{r=1}^R$, respectively, where $\alpha \in (0, 1)$ is the nominal size of the tests. Reject the null hypothesis \eqref{hypotheses_T1} or \eqref{hypotheses_T2} if $V_{jn}$ and $W_{jn}$ defined in~\eqref{stats_first}-\eqref{stats_last} are, respectively, larger than $\hat{q}_{V^{*}}(1-\alpha)$ or $\hat{q}_{W^{*}}(1-\alpha)$.
\end{enumerate}

The formulas in part~3 of the steps above are obtained by inserting estimated contact sets and resampled empirical processes in the place of population-level quantities into the functions shown in part~1 of Theorem~\ref{thm:obs_teststats}.  

The resampled statistics are calculated by imposing the null hypothesis and assuming that the region $\calX_0^j(P)$ is the only part of the domain that provides a nondegenerate contribution to the asymptotic distribution of the statistic under the null.  The two cases of each part in the maximum arise from trying to impose the null behavior on the resampled supremum norm statistics, even when it appears the null is violated based on the value of the sample statistic.  A simple alternative way to conduct inference would be to assume the least-favorable null hypothesis that $F_A \equiv F_B$, and to resample using all of $\calX$.  However, this may result in tests with lower power \citep{LintonSongWhang10}~--- power loss arises in situations where $\calX_0(P) \subset \calX$ (strictly), so that the $T_j$ process is only nondegenerate on a subset, while bootstrapped processes that assume $\calX_0(P) = \calX$ would look over all of $\calX$ and result in a stochastically larger bootstrap distribution than the true distribution.

The next result shows that our tests based on the resampling schemes described above have accurate size under the null hypothesis.  In order to metrize weak convergence we use test functions from the set $BL_1$, which denotes Lipschitz functions $\R \rightarrow \R$ that have constant 1 and are bounded by 1.

\begin{theorem} \label{thm:resample_consistent}
  Make assumptions~\ref{assumptionA_first}-\ref{assumptionA_last} and suppose that $P \in \calP_{00}$. %Let $\hat{q}_{V^*_j}(1-\alpha)$ and $\hat{q}_{W^*_j}(1-\alpha)$ be the $(1-\alpha)^{\text{th}}$ sample quantile from the bootstrap distributions as described in the routines above.
  Let $X$ denote the sample observations.  Then for $j = 1, 2$, the bootstrap is consistent: 
  \begin{equation*}
    \sup_{f \in BL_1} \left| \ex{f(V_{n}^*) | X} - \ex{f(V_1)} \right| = o_P(1)
  \end{equation*}
  and
  \begin{equation*}
    \sup_{f \in BL_1} \left| \ex{f(W_{n}^*) | X} - \ex{f(W_1)} \right| = o_P(1),
  \end{equation*}
  where $V_1$ and $W_1$ are defined in Theorem~\ref{thm:obs_teststats}.  In particular, when $P \in \calP_{00}$ the resampling procedure outlined above results in asymptotically valid inference: for any $P \in \calP_{00}$, letting $q_{V^*_j}(1 - \alpha) = \lim_{R \rightarrow \infty} \hat{q}_{V^*_j}(1 - \alpha)$ and $q_{W^*_j}(1 - \alpha) = \lim_{R \rightarrow \infty} \hat{q}_{W^*_j}(1 - \alpha)$,
  \begin{equation*}
    \limsup_{n \rightarrow \infty} \prob{ V_{jn} > q_{V_j^*}(1 - \alpha) } \leq \alpha
  \end{equation*}
  and
  \begin{equation*}
    \limsup_{n \rightarrow \infty} \prob{ W_{jn} > q_{W_j^*}(1 - \alpha) } \leq \alpha,
  \end{equation*}
  with equality when the distributions of $V_j$ and $W_j$ are strictly increasing at their $(1-\alpha)$-th quantiles.
\end{theorem}

The result in above theorem is stated in terms of the limiting variables $V_1$ and $W_1$ and bootstrap analogs.  $V_1$ and $W_1$, using the functional delta method, are Hadamard directional derivatives of a chain of maps from the marginal distribution functions $F$ to the real line, and the derivatives are most compactly expressed as the definitions in Theorem~\ref{thm:obs_teststats}.  

The bootstrap variables combine conventional resampling with finite-sample estimates of the maps defined in Part~1 of Theorem~\ref{thm:obs_teststats}, which is a resampling approach proposed in \citet{FangSantos19}.  Their result is actually more general~--- it states that with a more flexible estimator $V_n^*$, we would obtain bootstrap consistency for $P$ in the null and alternative regions.  Because our focus is on testing $F_A \succeq_{LASD} F_B$, however, our resampling scheme, and Theorem~\ref{thm:resample_consistent}, are done under the imposition of the null hypothesis.  The resampling consistency result in Theorem~\ref{thm:resample_consistent} implies that our bootstrap tests have asymptotically correct size for all probability distributions in the null region, in the same sense as was stressed in~\citet{LintonSongWhang10}.  A formal statement showing size control over all of $\calP_0$ is given in Theorem~\ref{thm:resample_size} in Appendix~\ref{app:hadamard}.  Along with Part 4 of Theorem~\ref{thm:obs_teststats}, Theorem~\ref{thm:resample_size} additionally implies that our tests are consistent, that is, that their power to detect violations from the null represented by fixed alternative distributions tends to one.  This is because the resampling scheme produces asymptotically bounded critical values, while the test statistics diverge under the alternative.

The behavior of bootstrap tests under the null and alternatives is most easily examined using distributions local to $P$.  We consider sequences of distributions $P_n$ local to the null distribution $P$ such that for a mean-zero, square-integrable function $\eta$, $P_n$ have distribution functions $F_n$ (where $P$ has CDF $F$) that satisfy
    \begin{equation} \label{local_alt_def}
      \lim_{n \rightarrow \infty} \int \left( \sqrt{n} \left( \sqrt{\dd F_n} - \sqrt{\dd F} \right) - \frac{1}{2} \eta \sqrt{\dd F} \right)^2 \rightarrow 0.
    \end{equation}
The behavior of the underlying empirical process under local alternatives satisfies Assumption 5 of \citet{FangSantos19} in a straightforward way \citep[Theorem 1]{Wellner92}.

\begin{theorem} \label{thm:resample_size}
  Make assumptions~\ref{assumptionA_first}-\ref{assumptionA_last} and suppose that $F_A \succeq_{LASD} F_B$.  Suppose that $\calX$ is convex.  Let $\hat{q}_{V^*_j}(1-\alpha)$ and $\hat{q}_{W^*_j}(1-\alpha)$ be the $(1-\alpha)^{\text{th}}$ sample quantile from the bootstrap distributions as described in the routines above, and let $q_{V_j^*}(1 - \alpha) = \lim_{R \rightarrow \infty} \hat{q}_{V_j^*}(1 - \alpha)$ and $q_{W_j^*}(1 - \alpha) = \lim_{R \rightarrow \infty} \hat{q}_{W_j^*}(1 - \alpha)$.  Then for $j = 1, 2$,
  \begin{enumerate}
    \item When $P \in \calP_0$ and $\{P_n\}$ satisfy~\eqref{local_alt_def} and $T_j(F_n)(x) \leq 0$ for all $x \geq 0$,
      \begin{equation*}
        \limsup_{n \rightarrow \infty} P_n \left\{ V_{jn} > q_{V^*_j}(1-\alpha) \right\} \leq \alpha 
      \end{equation*}
      and
      \begin{equation*}
        \limsup_{n \rightarrow \infty} P_n \left\{ W_{jn} > q_{W^*_j}(1-\alpha) \right\} \leq \alpha.
      \end{equation*}
    \item When $P \in \calP_{00}$ and $\{P_n\}$ satisfy~\eqref{local_alt_def} and $T_j(F_n)(x) \leq 0$ for all $x \geq 0$, and the distribution of $V$ or $W$ is increasing at its $(1-\alpha)^{\mathrm{th}}$ quantile,
      \begin{equation*}
        \lim_{n \rightarrow \infty} P_n \left\{ V_{jn} > q_{V^*_j}(1-\alpha) \right\} = \alpha 
      \end{equation*}
      and 
      \begin{equation*}
        \lim_{n \rightarrow \infty} P_n \left\{ W_{jn} > q_{W^*_j}(1-\alpha) \right\} = \alpha.
      \end{equation*}
  \end{enumerate}
\end{theorem}

In the Supplemental Appendix we provide Monte Carlo numerical evidence of the finite sample properties of both point- and partially-identified  methods. The simulations show that tests have empirical size close to the nominal, and high power against selected alternatives.

\section{Empirical illustration}\label{sec:applic} 

In this section we briefly illustrate the use of our approach using household-level data from a well-known experimental evaluation of alternative welfare programs in the state of Connecticut, documented in \citet{BitlerGelbachHoynes06}. Aid to Families with Dependent Children (AFDC) was one of the largest federal assistance programs in the United States between 1935 and 1996. It consisted of a means-tested income support scheme for low-income families with dependent children, administered at the state level, but funded at the federal level. Following criticism that this program discouraged labor market participation and perpetuated welfare dependency, the Clinton administration enacted the 1996 Personal Responsibility and Work Opportunity Reconciliation Act (PRWORA), requiring all US states to replace AFDC with a Temporary Assistance for Needy Families (TANF) program. TANF programs differed amongst US states and were all fundamentally different from AFDC: they included strict time limits for the receipt of benefits and, simultaneously, generous earnings disregard schemes to incentivise work. 

Under the policy framework of TANF, the state of Connecticut launched its own program, called Jobs First (JF) in 1996: this  included the strictest time limit and also the most generous earnings disregard of all the US states. Nonetheless, there was a transition period during which a policy experiment was conducted by the Manpower Demonstration and Research Corporation (MDRC). A random sample of approximately 5000 welfare applicants was randomly assigned to one of two groups: half of them were assigned to JF and faced its eligibility and program rules; the other half were randomly assigned to AFDC (the program that JF aimed to replace in the state of Connecticut), thereby facing AFDC eligibility and program rules. 

The MDRC experimental data include rounded data on quarterly income for a pre-program assignment period and also for a post-program assignment period, thereby allowing one to quantify and compare the income gains and losses experienced by the households that were randomly assigned to JF and ADFC\footnote{\citet{BitlerGelbachHoynes06} conduct a test comparing features of households before random assignment and find that they do not differ significantly in terms of observable characteristics.  We check additionally that the income distributions were the same before the experiment split households among the two policies.  We use a conventional two-sided Cram\'er-von Mises test for the equality of distributions.  The statistic was approximately $0.78$ and its p-value was $0.55$, implying that before the experiment, the distributions are indistinguishable.}   \citet{BitlerGelbachHoynes06} use these experimental data to compare the distribution of income between the beneficiaries of AFDC and JF. They find that while JF made the majority of individuals better-off, it also made a significant number of worse-off, especially after the JF time limit kicks in and becomes binding.\footnote{\citet{BitlerGelbachHoynes06} focus on quantile treatment effects (QTEs).  If QTEs were to be used as a measure of the impact on any individual household in a welfare comparison, it would require the assumption of rank invariance across potential outcome distributions, which would be quite strong.  Note that \citet{BitlerGelbachHoynes06} do not make this assumption.} In our simple empirical illustration we draw on \citet{BitlerGelbachHoynes06} and consider ``AFDC'' and ``JF'' as our alternative policies (equivalent to policies $A$ and $B$ in the previous sections). We illustrate our methods by constructing a LASD partial order to support a policy choice between these two programs\footnote{Although not directly relevant for our empirical illustration, it can be mentioned that the debate on the replacement of AFDC by TANF combined political economy concerns and also normative considerations about the appropriateness of policy-makers causing income losses to parts of the population. \citet{AlesinaGlaeserSacerdote01} use the AFDC as an empirical proxy for the generosity of the welfare state in the US and show that changes to this program had the potential to sway the electorate. At the same time, normative arguments supporting policy-makers' loss-aversion have also been put forth in this context. Peter Edelman, then a senior advisor to President Clinton, resigned in protest against this policy change, calling the replacement of ADFC by TANF a "crucial moral litmus test", as it risked causing important income losses to some households.} and comparing it with the partial ordering that would emerge if loss aversion were not taken into consideration using conventional first order stochastic dominance (FOSD).\footnote{Because assignment is random, we assume that the distribution functions of gains and losses under each policy, $F_{JF}$ and $F_{AFDC}$, are point-identified by the differences in incomes before and after random assignment.} Along the lines of \citet{BitlerGelbachHoynes06} we make this comparison separately for the time period up until the JF time limit for the receipt of welfare benefits becomes binding and for the period after that.

\subsection{JF vs AFDC: LASD ordering}

To make welfare decisions in terms of gains and losses, we use data on household income changes, i.e. the difference between households' income after exposure to the program (JF or AFDC) and before exposure to that program. We make this analysis separately for the period before the JF time limit become binding (TL) and for after that. We thus call pre-TL observations those that were made after random assignment to either of the policies (JF or AFDC) but before the time limit; we call post-TL observations those made after the JF time limit.  We summarize household income (for both policies and pre/post TL periods) by averaging income over all quarters in the relevant time span.\footnote {We explored alternative definitions of our outcome of interest such as using the final quarter within the time span; generally these led to the same results, so we will not show them  for the purpose of this simple   illustration. } Changes in household income due to the AFDC and JF policies were defined as the natural logarithm of the average household income in all post-policy quarters (either JF or AFDC) minus the natural log of the average pre-policy quarterly household income.  Thus, our analysis applies LASD to these changes in two separate periods, the pre-TL period and the post-TL one.

The left-hand side of Table~\ref{tab:pointIDtests} shows the results of formal tests of the hypothesis~(\ref{hypotheses_general}) using $W_{2n}$ statistics (Cram\'er-von Mises statistics applied to the empirical $T_2$ process).\footnote{Results for the other test statistics are qualitatively the same. They are collected in an Online Supplemental Appendix.} For the pre-time limit period we cannot reject the hypothesis that $F_{JF} \succeq_{LASD} F_{AFDC}$.\footnote{For this time period we cannot even reject the null of equality in the distributions of changes in income between households assigned to JF and AFDC.} However, everything changes when we make this comparison taking into account the post-time limit period. As mentioned above, after the time limit becomes binding, \citet{BitlerGelbachHoynes06} show that a sizeable number of households in JF experience total income losses, as they stop receiving welfare transfers; this does not happen amongst households on ADFC, which does not have a time limit. In order to rank the distribution of income changes under JF and AFDC using LASD we test the hypothesis that $F_{JF} \succeq_{LASD} F_{AFDC}$. As shown in the left-hand side of Table 1, this hypothesis is rejected for every significance level, reflecting the greater weight placed on the income losses experienced by JF beneficiaries.

\begin{small}
\input{"table_of_tests.tex"}
\end{small}

\begin{figure}[ht!]
  {\centering \includegraphics[width=\columnwidth]{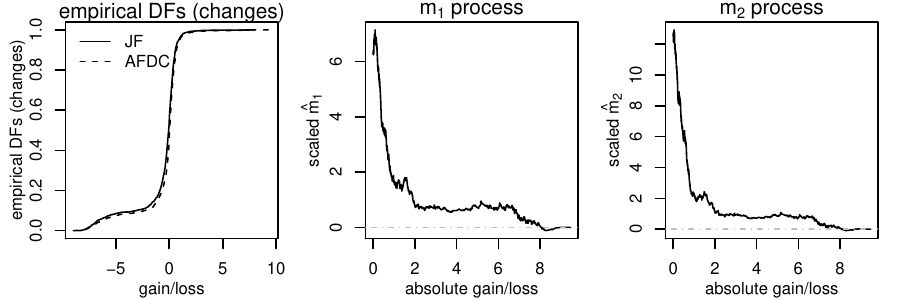}}
  \caption{Empirical distribution functions of changes in income near the JF time limit and the $T_{2}(F)$ coordinate processes that are used to test loss aversion-sensitive dominance.  The second and third panels show the two coordinate functions in $T_2(\bbF_n)$ defined in equation~\eqref{t2def}.  The large positive values in the second panel drive the rejection of the hypothesis $F_{JF} \succeq F_{AFDC}$ seen in Table~\ref{tab:pointIDtests}.} \label{fig:LASD}
\end{figure}

To investigate how this rejection occurs, Figure~\ref{fig:LASD} displays the CDFs of gains and losses under the AFDC and JF policies around the JF time limit, then the way that the two $T_2$ coordinate processes compare them~--- when looking at the coordinates in equation~\eqref{t2def}, large positive values correspond to a rejection of the hypothesis $F_{JF} \succeq_{LASD} F_{AFDC}$.  The positive parts of the $m_1$ and $m_2$ functions illustrated in the middle and right-hand plots of Figure~\ref{fig:LASD} are squared and integrated over estimated contact sets to arrive at the test statistic in the lower left of Table~\ref{tab:pointIDtests}.  It can be seen in the second and third panels that the presumable reason that the JF policy does not dominate the AFDC policy using LASD is because the distribution of small gains and losses is more appealing in the AFDC program and the relation between small gains and small losses is preferable to JF.

\subsection{JF vs AFDC: first order stochastic dominance ordering}
What difference would it make if loss-aversion had been left out of this welfare ordering of social policies? In order to address this question we compare the welfare ordering obtained in the previous section with that obtained by ordering JF and AFDC according to first order stochastic dominance (FOSD). Using FOSD, the only relevant comparison is between the post-policy household income under JF and AFDC (household income before exposure to these policies is not material). We thus define our outcome of interest in levels, i.e. the natural log of the average household income under JF or AFDC. As before, we do this analysis separately for the two relevant time periods: before the JF time limit becomes binding and after it does. 

The right-hand side of Table 1 shows the  result of our FOSD tests. Either way post-policy outcomes are measured, we cannot reject the null that $G_{JF} \succeq_{FOSD} G_{AFDC}$. When measurements are made before and after exposure to the policy this is unsurprising, as prior to the JF time limit becoming binding none of the policies produces large income losses. However, even after the JF time limit becomes binding, the FOSD test still does not allow us to reject $G_{JF} \succeq_{FOSD} G_{AFDC}$, while the LASD test would lead us to categorically reject the dominance of JF over AFDC. This simple empirical illustration shows that, in practice, the consideration of loss aversion can change the welfare ordering of social policies.

\begin{figure}[ht!] 
  \begin{center}
  \includegraphics[width=0.75\columnwidth]{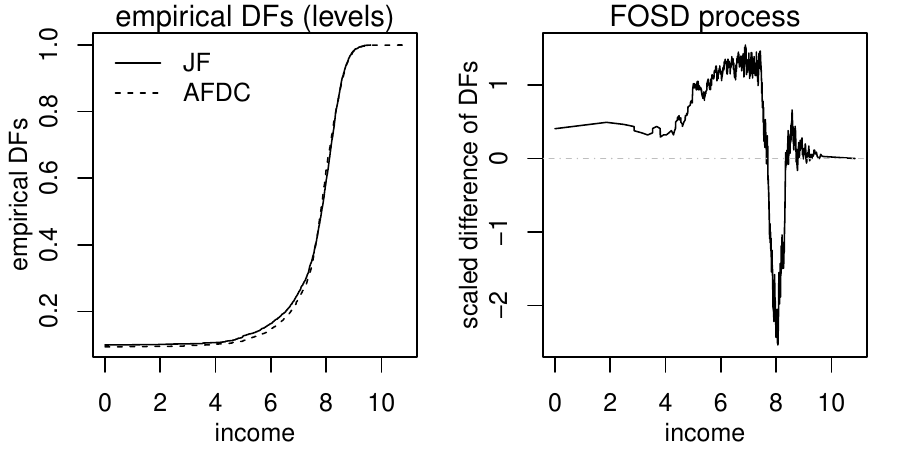}
  \end{center}
  \caption{Empirical distribution functions of (levels of) post-TL income and the resulting process that is used to test first-order stochastic dominance.  Although there are positive values in the second panel, they are not large enough to drive a rejection of the hypothesis $G_{JF} \succeq_{FOSD} G_{AFDC}$, as seen in Table~\ref{tab:pointIDtests}.} \label{fig:FOSD}
\end{figure}

Figure~\ref{fig:FOSD} shows an analogous investigation into the way analysis would typically be conducted using first order stochastic dominance to compare outcomes, using contact sets as in~\citet{LintonSongWhang10}.  The contact set was estimated using $a_n = 4\log(\log(n))$, corresponding to the tuning parameter choice of that paper, for both the FOSD and LASD tests (the smaller sequence $c_n = \sqrt{\log(\log(n))}$ was used for estimating near-maximizing sets in LASD tests).  The left-hand plot in the figure shows the empirical distribution functions of outcomes under each program after the JF time limit.  That is, the functions are based on levels of income rather than changes in income.  The scaled difference $\sqrt{n}(\bbG_{n,JF} - \bbG_{n,AFDC})$ is displayed as the criterion function in the right-hand side of the panel, and the square of the positive part of this function is integrated over an estimated contact set.  As the lower-right test in Table~\ref{tab:pointIDtests} indicates, these differences are sometimes mildly positive, so that $G_{AFDC}$ is occasionally below $G_{JF}$ (especially at lower income levels) but the difference is not large enough to indicate a rejection of the hypothesis that $G_{JF} \succeq_{FOSD} G_{AFDC}$, as indicated by the p-value of the test.  Because outcomes are measured in levels, there is no way to measure whether they represent gains or losses for agents, and so a simple difference is used here instead of the comparison that accounts for loss aversion used with changes in income.

%\section{Discussion}\label{sec:conclusion}
\section{Conclusion}\label{sec:conclusion}

Public policies often result in gains for some individuals and losses for others. 
We define a social preference relation for distributions of gains and losses caused by a policy: loss aversion-sensitive dominance (LASD).  We relate these social preferences to criteria that depend solely on distribution functions.  The assumption of loss aversion can lead to a welfare ranking of policies that is different from the one that would be brought about if classic utility theory and first-order stochastic dominance were used. We then propose empirically testable conditions for LASD based on our CDF-based criterion functions. Because data may come as differences between underlying random variables, we propose a point-identified version of these conditions and also a partially identified analog.
We develop inference methods to formally test LASD relations and derive the corresponding statistical properties. 
We show that  resampling techniques, tailored to specific features of the criterion functions, can be used to conduct inference. Finally, : illustrate  our LASD criterion and inference methods with a simple empirical application that uses data from a well known evaluation of a large income support policy in the US. This shows that the ranking of policy options depends crucially on whether changes or levels are used and whether or not one takes individual loss aversion into account.

\newpage
\appendix
\noindent \begin{Large}\textbf{Appendix}\end{Large} 
\vspace{-0.5cm}
\section{Proof of results}\label{app:proofs}

\subsection{Results in Section \ref{sec:basic}}

\begin{proof}[Proof of Proposition~\ref{prop:svf_alt}]
  Equation~\eqref{eq:svf} implies that
  \begin{equation} \label{svf_twopart}
    W(F) = \int_{\mathbb{R}_-} v(x) \dd F(x) + \int_{\mathbb{R}_+} v(x) \dd F(x).
  \end{equation}
  We will now re-write the two parts of ~\eqref{svf_twopart} using integration by parts,  normalization $v(0) = 0$ and the fact that $F$ has bounded support i.e. there exist $R_1 > 0$ and $R_2 < 0$ such that $\int_{R_2}^{R_1} f(x) \dd x = 1$, in which case we have $1-F(R_2)=0$ and $F(R_1)$=0.  For the first part of~(\ref{svf_twopart}) we obtain 
  \begin{align*}
    \int_{\mathbb{R}_-} v(x) \dd F(x) &=  \lim_{R \rightarrow -\infty} \int_R^0 v(x)\dd F(x) \\
    {} &= \lim_{R \rightarrow -\infty} \sbr{v(x)F(x)|_R^0 - \int_R^0 v'(x)F(x)\dd x} \\
    {} &= -\int_{-\infty}^0 v'(x)F(x)\dd x
  \end{align*}
  and for the second part~(\ref{svf_twopart}) we have 
  \begin{align*}
    \int_{\mathbb{R}_+} v(x) \dd F(x) &= - \int_{\mathbb{R}_+} v(x) \dd (1-F)(x) \\
    {} &= - \lim_{R \rightarrow \infty} \int_0^R v(x) \dd (1-F)(x) \\
    {} & = - \lim_{R \rightarrow \infty} \sbr{v(x)(1-F(x))|_0^R - \int_0^R v'(x) (1-F(x))\dd x} \\
    {} &= \int_0^{\infty} v'(x) (1-F(x))\dd x.
  \end{align*}
  Putting these two parts together yields~(\ref{svf2}).
\end{proof}

\subsection{Proofs of results in Section \ref{sec:ptsd}}

\begin{proof}[Proof of Theorem \ref{thm:ptfsd}] 
Notice that \eqref{eq:ptfsd} is equivalent to both \eqref{eq:ptfsd1} and \eqref{eq:ptfsd2}; in this proof we use the latter two conditions.  Using Proposition~\ref{prop:svf_alt} we rewrite $W(F_{A}) \geq W(F_{B})$ as the equivalent condition
\[
-\int_{-\infty}^0 v'(z)F_{A}(z)\dd z + \int_0^{\infty} v'(z)(1-F_{A}(z))\dd z \geq -\int_{-\infty}^0 v'(z)F_{B}(z)\dd z + \int_0^{\infty} v'(z)(1-F_{B}(z))\dd z.
\]

Rearranging terms we find this is equivalent to
\[
 \int_{-\infty}^0 v'(z)F_{B}(z)\dd z - \int_{-\infty}^0 v'(z)F_{A}(z)\dd z \geq \int_0^{\infty} v'(z)(1-F_{B}(z))\dd z -\int_0^{\infty} v'(z)(1-F_{A}(z))\dd z
\]
or simply
\[
 \int_{-\infty}^0 v'(z)(F_{B}(z)-F_{A}(z))\dd z \geq \int_0^{\infty} v'(z)(F_{A}(z)-F_{B}(z))\dd z.
\]
This is in turn equivalent to
\[
\int_0^{\infty} v'(-z)(F_{B}(-z)-F_{A}(-z))\dd z \geq \int_0^{\infty} v'(z)(F_{A}(z) - F_{B}(z)))\dd z
\]
or
\[
-\int_0^{\infty} v'(-z)(F_{A}(-z)-F_{B}(-z))\dd z \geq \int_0^{\infty} v'(z)(F_{A}(z) - F_{B}(z)))\dd z.
\]
Adding $v'(z) (F_{A}(-z)-F_{B}(-z))$ to both sides we find this is equivalent to
\begin{equation}        \label{ineq1}
\int_0^{\infty} (v'(z)-v'(-z))(F_{A}(-z)-F_{B}(-z))\dd z \geq \int_0^{\infty} v'(z)(F_{A}(z) - F_{B}(z) + F_{A}(-z) - F_{B}(-z))\dd z.
\end{equation}
Utilizing the assumptions of loss aversion and non-decreasingness given in Definition~\ref{deff:properties}, \eqref{eq:ptfsd1} and \eqref{eq:ptfsd2} are sufficient for \eqref{ineq1} to hold for any $v$. Condition \eqref{eq:ptfsd2} is due to the fact that
\begin{align*}
  F_{A}(x)-F_{B}(x)+F_{A}(-x)-F_{B}(-x) &\leq 0 \quad \forall x\geq0 \\
  \intertext{is equivalent to the condition}
  1 - F_{A}(x) - F_{A}(-x) &\geq 1 - F_{B}(x) - F_{B}(-x) \quad \forall x\geq0.
\end{align*}

We now show that conditions \eqref{eq:ptfsd1} and \eqref{eq:ptfsd2} are also necessary by means of a contradiction to \eqref{ineq1}. To this end, assume that there exists some $x>0$ such that $F_{A}(-x)-F_{B}(-x) > 0$. From the fact that the distribution function is right continuous, it follows that there is a neighbourhood $(a,b)$, $b>a > 0$, such that for all $x \in (a,b)$, $F_{A}(-x)-F_{B}(-x) > 0$. For arbitrarily small $\epsilon > 0$, consider the value function
\[
v_1(x)=\begin{cases}
a-b & x\leq -b - \epsilon \\
\frac{1}{4 \epsilon} (x + b + \epsilon)^2 + a -b & x \in (-b - \epsilon, -b + \epsilon) \\
x+a & x\in(-b+ \epsilon ,-a - \epsilon) \\
\frac{1}{4 \epsilon} (x + a - \epsilon)^2 & x \in (-a - \epsilon, -a + \epsilon) \\
0 & x\geq -a + \epsilon.
\end{cases}
\]
Note that this $v_1$ satisfies  Definition~\ref{deff:properties}.  Further, for $x \in (a,b)$, $v_1'(-x) > 0 = v_1'(x)$. Therefore
\[
\int_0^{\infty} (v_1'(z)-v_1'(-z))(F_{A}(-z)-F_{B}(-z))\dd z < 0,
\]
while
\[
\int_0^{\infty} v_1'(z)(F_{A}(z) - F_{B}(z) + F_{A}(-z) - F_{B}(-z))\dd z = 0,
\]
because $v_1'(x) = 0$ for $x \geq 0$. This contradicts \eqref{ineq1}.

The second condition can be proven similarly. Assume that there exists a neighbourhood $(a,b)$, $0 < a < b$ such that for all $x \in (a,b)$, $\rbr{1-F_{A}(x)}-F_{A}(-x)< \rbr{1-F_{B}(x)}-F_{B}(-x)$. Take $v_2(x) = \text{sgn}(x) \times v_1 (x)$. Using $v_2$ we find
\[
\int_0^{\infty} (v_2'(z)-v_2'(-z))(F_{A}(-z)-F_{B}(-z))\dd z = 0 
\]
while
\[ 
\int_0^{\infty} v_2'(z)(F_{A}(z) - F_{B}(z) + F_{A}(-z) - F_{B}(-z))\dd z > 0,
\]
which is a contradiction.

\end{proof}

%%%%%%%%%%%%%%%%%%%%%%%%%%%%%%%%%%%%%%%%%%%%%%%%%%%%%%%%%%%%%%%%%%%%%%%%%%%%%%%%%%%%%%%%%%%%%%%%%%%%%%%%%%%%%%%%%%%%%%%%%%%%%%%%%%%%%%%%%%%%%%%%%%%%%%%%%%%%%%%%%%%%%%%%%%%%%%%%%%%%
%%%%%%%%%%%%%%%%%%%%%%%%%%%%%%%%%%%%%%%%%%%%%%%%%%%%%%%%%%%%%%%%%%%%%%%%%%%%%%%%%%%%%%%%%%%%%%%%%%%%%%%%%%%%%%%%%%%%%%%%%%%%%%%%%%%%%%%%%%%%%%%%%%%%%%%%%%%%%%%%%%%%%%%%%%%%%%%%%%%%

%%%%%%%%%%%%%%%%%%%%%%%%%%%%%%%%%%%%%%%%%%%%%%%%%%%%%%%%%%%%%%%%%%%%%%%%%%%%%%%%%%%%%%%%%%%%%%%%%%%%%%%%%%%%%%%%%%%%%%%%%%%%%%%%%%%%%%%%%%%%%%%%%%%%%%%%%%%%%%%%%%%%%%%%%%%%%%%%%%%%
%%%%%%%%%%%%%%%%%%%%%%%%%%%%%%%%%%%%%%%%%%%%%%%%%%%%%%%%%%%%%%%%%%%%%%%%%%%%%%%%%%%%%%%%%%%%%%%%%%%%%%%%%%%%%%%%%%%%%%%%%%%%%%%%%%%%%%%%%%%%%%%%%%%%%%%%%%%%%%%%%%%%%%%%%%%%%%%%%%%%

\begin{proof}[Proof of Corollary \ref{corr:statusquo}]
	We first notice that $F_{A} \succeq_{FOSD} F_{SQ}$ is equivalent to the event
	\begin{equation}\label{eq:positiveSupport}
	    \left\{ F_A \text{ is supported on }  \mathbb{R}_+ \right\}.
	\end{equation}
	Property \eqref{eq:positiveSupport} easily implies that $F_{A} \succeq_{LASD} F_{SQ}$, which follows by Property 1 of Definition~\ref{deff:properties}. On the other hand  one checks that   $$
	    v(x):=\begin{cases}
	x & x\leq0\\
	0 & x>0
	\end{cases}
	    $$
	    fulfills Definition \ref{deff:properties}. Thus  $F_{A} \succeq_{LASD} F_{SQ}$ implies \eqref{eq:positiveSupport}.
\end{proof}

\begin{proof}[Proof of Theorem~\ref{thm:boundsfsd}] 
	Given the bounds inequality, we have
\[
L_{B}(-x)-U_{A}(-x)\leq F_{B}(-x)-F_{A}(-x) \leq U_{B}(-x)-L_{A}(-x)
\]
and
\[
L_{A}(x)-U_{B}(x)\leq F_{A}(x) - F_{B}(x) \leq U_{A}(x) - L_{B}(x),
\]
from which it is clear that \eqref{ptfsdsuff} is a sufficient condition. As a necessary condition we have \eqref{ptfsdnec} because using \eqref{eq:ptfsd} we have
\[
U_B(-x)-L_A(-x)\geq F_B(-x)-F_A(-x)\geq \max\{0, F_A(x)-F_B(x)\}\geq \max\{0, L_A(x)-L_B(x)\}.
\]
\end{proof}

%%%%%%%%%%%%%%%%%%%%%%%%%%%%%%%%%%%%%%%%%%%%%%%%%%%%%%%%%%%%%%%%%%%%%%%%%%%%%%%%%%%%%%%%%%%%%%%%%%%%%%%%%%%%%%%%%%%%%%%%%%%%%%%%%%%%%%%%%%%%%%%%%%%%%%%%%%%%%%%%%%%%%%%%%%%%%%%%%%%%
%%%%%%%%%%%%%%%%%%%%%%%%%%%%%%%%%%%%%%%%%%%%%%%%%%%%%%%%%%%%%%%%%%%%%%%%%%%%%%%%%%%%%%%%%%%%%%%%%%%%%%%%%%%%%%%%%%%%%%%%%%%%%%%%%%%%%%%%%%%%%%%%%%%%%%%%%%%%%%%%%%%%%%%%%%%%%%%%%%%%

%\begin{proof}[Proof of Theorem \ref{thm:boundsssd}]
%	This follows from \eqref{eq:ptssd3} and the inequality bounds:
%	\[\int_{-y}^x (L_{A}(z)-U_{B}(z))\dd z \leq \int_{-y}^x (F_{A}(z)-F_{B}(z))\dd z \leq \int_{-y}^x (U_{A}(z)-L_{B}(z))\dd z. \]
% \end{proof}                       

%%%%%%%%%%%%%%%%%%%%%%%%%%%%%%%%%%%%%%%%%%%%%%%%%%%%%%%%%%%%%%%%%%%%%%%%%%%%%%%%%%%%%%%%%%%%%%%%%%%%%%%%%%%%%%%%%%%%%%%%%%%%%%%%%%%%%%%%%%%%%%%%%%%%%%%%%%%%%%%%%%%%%%%%%%%%%%%%%%%%
%%%%%%%%%%%%%%%%%%%%%%%%%%%%%%%%%%%%%%%%%%%%%%%%%%%%%%%%%%%%%%%%%%%%%%%%%%%%%%%%%%%%%%%%%%%%%%%%%%%%%%%%%%%%%%%%%%%%%%%%%%%%%%%%%%%%%%%%%%%%%%%%%%%%%%%%%%%%%%%%%%%%%%%%%%%%%%%%%%%%

%%%%%%%%%%%%%%%%%%%%%%%%%%%%%%%%%%%%%%%%%%%%%%%%%%%%%%%%%%%%%%%%%%%%%%%%%%%%%%%%%%%%%%%%%%%%%%%%%%%%%%%%%%%%%%%%%%%%%%%%%%%%%%%%%%%%%%%%%%%%%%%%%%%%%%%%%%%%%%%%%%%%%%%%%%%%%%%%%%%%
%%%%%%%%%%%%%%%%%%%%%%%%%%%%%%%%%%%%%%%%%%%%%%%%%%%%%%%%%%%%%%%%%%%%%%%%%%%%%%%%%%%%%%%%%%%%%%%%%%%%%%%%%%%%%%%%%%%%%%%%%%%%%%%%%%%%%%%%%%%%%%%%%%%%%%%%%%%%%%%%%%%%%%%%%%%%%%%%%%%%
 
\begin{proof}[Proof of Corollary~\ref{corr:boundsfsd}]
Recall Corollary~\ref{corr:statusquo} implied that when $F_B$ is a status quo distribution, the FOSD and LASD relations are equivalent.  Then $F_{A} \succeq_{FOSD} F_{SQ}$ implies that $F_A(-x) = 0$ for all $x \geq 0$ because $F_{SQ}(-x) = 0$ for all $x \geq 0$.  Therefore a sufficient condition for $F_{A} \succeq_{LASD} F_{SQ}$ is that $U_A(-x) = 0$ for all $x \geq 0$.  Similarly, if $F_A \succeq_{LASD} F_{SQ}$, equivalent to $F_A \succeq_{FOSD} F_{SQ}$, then it must be the case that $F_A(-x) = 0$ for all $x \geq 0$, implying that $L_A(-x) = 0$ as well.
\end{proof}

\subsection{Results in Section \ref{sec:inference}}

\begin{proof}[Proof of Theorem~\ref{thm:obs_teststats}]
  For Part 1 note that if $\calP \in \calP_{00}$ then by definition, $\calX_0^k(P) \neq \ns$ for some $k \in \{1, 2\}$ and for all $x \in \calX_0^k(P)$, $m_k(x) = 0$.  For any $k$ such that $\calX_0^k(P) \neq \ns$, that set is the limit of $\calM_k(\epsilon)$ in the Painlev\'e-Kurotowski sense as $\epsilon \searrow 0$, because $\calM_k(\epsilon)$ form a monotone sequence of sets and Exercise 4.3 of \citet{RockafellarWets98} implies the limit is $\calX_0^k(P)$. Then the supremum is achieved and $\lim_{\epsilon \searrow 0} \calM_k(\epsilon) = \calX_0^k(P)$ (in the Painlev\'e-Kuratowski sense) for at least one coordinate, so that suprema are taken over at least one of $\calX_0^1(P)$ and $\calX_0^2(P)$ and whichever coordinate satisfies this condition will contribute to the asymptotic distribution.  Note that for all $x \in \calX_0(P)$, $\sqrt{n} T_1(\bbF_n)(x) = \sqrt{n}(T_1(\bbF_n) - T_1(F))(x)$.  Lemma~\ref{lem:t1diff} and the null hypothesis, which implies $\calX_0^k(P) \neq \ns$ for $k \in \{1, 2\}$, imply the result for $V_1$ and $W_1$.  

  To show Part 2, note that $T_2$ is a linear map of $F$, and assuming that $\calX_0^k(P) \neq \ns$ for $k \in \{1, 2\}$, we have that its weak limit (for whichever set is nonempty) is $\sup_{x \in \calX_0^k(P)} (T_{2k}(\calG_F)(x))^+$ by Lemma~\ref{lem:t1diff}.  Breaking $\calX_0(P)$ into its two subsets and assuming the null hypothesis is true results in the same behavior as the supremum norm statistic from the first part (using the definition of the supremum norm in two coordinates as the maximum of the two suprema).  The same reasoning holds for the $L_2$ statistic in Part 2.

  Part 3 follows from the behavior of the test statistics over $\{x \in \calX : m_1(x) < 0, m_2(x) < 0\}$ described in Lemma~\ref{lem:t1diff}.  To show Part 4 for $V_{1n}$ suppose that for some $x^*$, $T_1(F)(x^*) = \xi > 0$.  Then $\sup_{x \in \calX} \sqrt{n} T_1(\bbF_n)(x) \geq \sqrt{n} (T_1(\bbF_n)(x^*) - T_1(F)(x^*)) + \sqrt{n} \xi$.  Then
  \begin{multline*}
    \liminf_{n \rightarrow \infty} \prob{ \sup_{x \geq 0} \sqrt{n} T_1(\bbF_n)(x) > c} \\
    \geq \lim_{n \rightarrow \infty} \prob{ \sqrt{n}(T_1(\bbF_n)(x^*) - T_1(F)(x^*)) > c - \sqrt{n} \xi} \rightarrow 1,
  \end{multline*}
  where the last convergence follows from the delta method applied to $\sqrt{n}(\bbF_n(x^*) - F(x^*))$, which converges in distribution to a tight random variable.  The proof for the other statistics is analogous.
\end{proof}

\begin{proof}[Proof of Theorem~\ref{thm:resample_consistent}]
  This theorem is an application of Theorems 3.2 and 3.3 of \citet{FangSantos19}.  Define the statistics $V_1$ and $W_1$ as maps from $F$ to the real line using $\nu$ and $\omega$ defined in equations~\eqref{nu_def} and~\eqref{omega_def} in Lemma~\ref{lem:t1diff}, and let their estimators be defined as in part 3 of the resampling scheme.  Their Assumptions 1-3 are satisfied either by the definitions of $\nu$ and $\omega$ and Lemma~\ref{lem:t1diff}, the standard convergence result $\sqrt{n}(\bbF_n - F) \cw \calG_F$ \citep[Theorem 2.8.4]{vanderVaartWellner96} and the choice of bootstrap weights.  We need to show that their Assumption 4 is also satisfied.  Write either function as $\|h_1^+\| + \| h_1^+ \vee h_2^+ \| + \|h_2^+\|$ using the desired norm.  Both norms satisfy a reverse triangle inequality, and using the fact that $|(x)^+ - (y)^+| \leq |x - y|$, the difference for two functions $g$ and $h$ is bounded by $\|g_1 - h_1\| + \| g_1 \vee g_2 - h_1 \vee h_2 \| + \|g_2 - h_2\|$.  The first difference is bounded by $2\|g - h\|$, and the second and the third are bounded by $4\|g - h\|$.  Rewriting equations~\eqref{vstar_def} and~\eqref{wstar_def} as functionals of differential directions $h$, define
  \begin{equation*}
	\hat{\nu}_n'(h) = \begin{cases} 
	\left( \max_{x \in \hat{\calM}^{\hat{k}}} h_{\hat{k}}(x) \right)^+ & | \max \hat{m}_{1n} - \max \hat{m}_{2n} | > c_n \\
	\max \left\{ 0, \max_{x \in \hat{\calM}^1} h_1(x), \max_{x \in \hat{\calM}^2} h_2(x) \right\} & | \max \hat{m}_{1n} - \max \hat{m}_{2n} | \leq c_n
	\end{cases}
\end{equation*}
and
\begin{equation}
	\hat{\omega}_n'(h) = \left( \int_{\hat{\calX}_0^1} \left( \left( h_1(x) \right)^+ \right)^2 \dd x + \int_{\hat{\calX}_0^2} \left( \left( h_2(x) \right)^+ \right)^2 \dd x \right)^{1/2}.
\end{equation}
  
  Because both $\nu$ and $\omega$ are Lipschitz, Lemma~S.3.6 of \citet{FangSantos19} implies we need only check that $|\hat{\nu}'_n(h) - \nu_F'(h)| = o_P(1)$ and $|\hat{\omega}'_n(h) - \omega_F'(h)| = o_P(1)$ for each fixed $h$.  This follows from the consistency of the contact set and $\epsilon$-argmax estimators.  The consistency of these estimators follow from the uniform law of large numbers for the $\epsilon$-maximizing sets, and the tightness of the limit $\calG_F$ for the contact sets, which implies that $\lim_n \prob{\sqrt{n}\|\bbF_n - F\|_\infty \leq \sqrt{n}a_n} = 1$.
  \end{proof}

\newpage
\singlespace
\bibliographystyle{econometrica}
\bibliography{distrwelfaretest}

\newpage
\input{"Supplemental_Appendix.tex"}

%\includepdf[pages=-]{Supplemental_Appendix.pdf}

\end{document}

%% file: table_of_tests.tex
%latex.default(tab, file = "table_of_tests.tex", label = "tab:pointIDtests",     rowname = timepv, title = "", ctable = TRUE, math.col.names = TRUE,     col.just = rep("c", 3), cgroup = c("LASD", "FOSD"), n.cgroup = c(1,         1), colheads = hypname, caption.loc = "bottom", caption = "Tests for inferring whether the Jobs First (JF) program would\n      be preferred to the Aid to Families with Dependent Children (AFDC).\n      Column titles paraphrase the null hypotheses in the tests.  The first \n      column uses changes in income and the second column measures income in \n      levels without regard to pre-policy income.  1999 bootstrap repetitions \n      used in each test.")%
\ctable[botcap,caption={Tests for inferring whether the Jobs First (JF) program would
      be preferred to the Aid to Families with Dependent Children (AFDC).
      Column titles paraphrase the null hypotheses in the tests.  The first 
      column uses changes in income and the second column measures income in 
      levels without regard to pre-policy income.  1999 bootstrap repetitions 
      used in each test.},label=tab:pointIDtests,pos=!tbp,]{lccc}{}{\FL
\multicolumn{1}{l}{\bfseries }&\multicolumn{1}{c}{\bfseries LASD}&\multicolumn{1}{c}{\bfseries }&\multicolumn{1}{c}{\bfseries FOSD}\NN
\cline{2-2} \cline{4-4}
\multicolumn{1}{l}{}&\multicolumn{1}{c}{$F_{JF} \succeq F_{AFDC}$}&\multicolumn{1}{c}{}&\multicolumn{1}{c}{$G_{JF} \succeq G_{AFDC}$}\ML
Before JF time limit&$0.1790$&&$0.2240$\NN
p-value&$0.9095$&&$0.8634$\NN
After JF time limit&$9.1380$&&$2.1285$\NN
p-value&$0.0000$&&$0.1436$\LL
}

%% file: Supplemental_Appendix.tex
\section*{Online Supplemental Appendix to ``Loss aversion and the welfare ranking of social policies''}

This supplement appendix contains numerical Monte Carlo simulations studying the empirical size and power of the statistical methods proposed in the main text and additional results for the empirical application in Section 5 of the main text.

\section{Monte Carlo simulations}

In this section, we compare the finite sample performances tests proposed in the text for testing the LASD null hypothesis. We describe the results of simulation experiments used to investigate the size and power properties of the tests described in the main text.  There are three simulation settings: a normal location model and a triangular model under point identification, and a normal location model under partial identification.

\begin{remark}[A note on computing point-identified criterion functions] \label{rem:compute}
Standard empirical distribution functions are used to estimate the marginal distributions $F_A$ and $F_B$.  However, the definitions of the $T_1$ and $T_2$ criterion functions contain $F_k(-x)$ terms, making the plug-in $T_j(\bbF_n)$ left-continuous at some sample observations.  Therefore some care must be taken when evaluating them because there may be regions that are relevant for evaluation (i.e., the location of the supremum) that are not attained by any sample observations.  This could be dealt with approximately by evaluating the functions on a grid.  Instead, we evaluate the function approximately at all the points where it changes its value.  For example, let $X_n$ denote the pooled sample (of size $(n_A + n_B)$) of $X_A$ and $X_B$ observations.  Then we evaluate $T_j$ at the points $\tilde{X}_n = 0 \cup X_n^+ \cup \{ X_n - \epsilon\}^-$, where $X_n^+$ and $X_n^-$ refer to the positive- and negative-valued elements of the pooled sample $X_n$ and $\epsilon$ is a very small amount added to each element of $X_n$, for example, the square root of the machine's double-precision accuracy.  When evaluating the $L_2$ integrals from an observed sample, the domain can be set to $[0, \tilde{x}_{max}]$, where $\tilde{x}_{max}$ is the largest point in the evaluation set $\tilde{X}_n$, because the integrand is identically zero above that point.
\end{remark}

\subsection{Normal model, identified case}
In this experiment there are two independent, Gaussian random variables that represent point-identified outcomes.  The scale of both distributions is set to unity, the location of distribution $A$ is set to zero and the location of distribution $B$ is allowed to vary.  Letting $\mu_B$ denote the location of distribution $B$, tests should not reject the null $H_0: F_A \succeq_{LASD} F_B$ when $\mu_B \leq 0$ and should reject the null when $\mu_B > 0$.  This is a case where $\calP_{00}$ is a singleton, which is when $\mu_B = 0$.

We select constant sequences using this model in a preliminary round of simulation (available in an online repository).  Let $n = n_A + n_B$.  The estimated contact sets $\hat{\calX}_0^k = \{x \in \calX : |\hat{m}_{kn}(x)| \leq a_n\}$ worked well using $a_n = 4 \log(\log(n)) / \sqrt{n}$.  For estimated $\epsilon$-maximizer sets $\hat{\calM}^k = \{x \in \calX : \hat{m}_{kn}(x) > \sup \hat{m}_{kn}(x) - b_n\}$ we used $b_n = \sqrt{\log(\log(n)) / n}$.  For deciding on which coordinate appeared significantly larger than the other, or whether both coordinates reached approximately the same supremum, that is, when estimating $|\max \hat{m}_{1n}(x) - \max \hat{m}_{2n}(x)| \leq c_n$,  we used the same constant sequence as $b_n$, that is, $c_n = \sqrt{\log(\log(n)) / n}$.  These sequences were used after preliminary simulations with the normal model, and were used in the other two simulations as well (with $n = n_0 + n_A + n_B$ in the partially-identified setting).

\begin{figure}[ht] 
  \centering
  \includegraphics[width = 0.9\columnwidth]{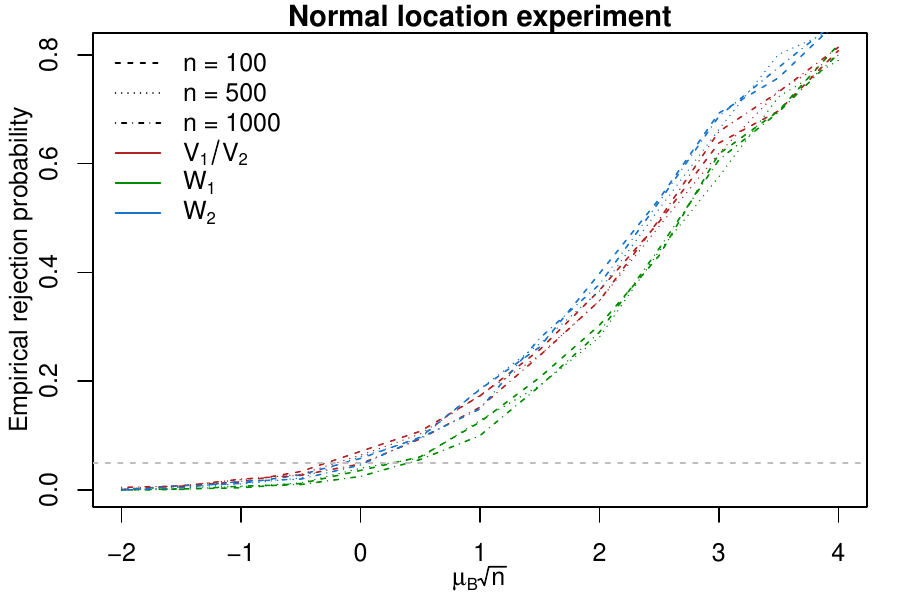}
  \caption{Empirical rejection probabilities of the LASD tests in the point identified normal location model experiment.  The tests are of nominal 5\% size, should have exactly 5\% rejection probability when $\mu_B = 0$ and should reject when $\mu_B > 0$.  $V_{1n}$ and $V_{2n}$ tests have identical behavior so only $V_{1n}$ results are shown.  Samples of sizes 100, 500 and 1000 correspond respectively to 499, 999 and 1999 bootstrap repetitions.  Distributions are local to the boundary of the null region, which is where $\mu_B = 0$.  1000 simulation repetitions.} \label{fig:normal_pointID}
\end{figure}

The size and power of the tests is good in this example, as can be seen in Figure~\ref{fig:normal_pointID}.  The mean of distribution $B$ ran from $-2 / \sqrt{n}$ to $4 / \sqrt{n}$ so the alternatives are local to the boundary of the null region.  Sample sizes were identical for both samples and set equal to 100, 500 or 1,000.  When resampling, the number of bootstrap repetitions was set equal to 499 (for samples of size 100), 999 (for samples of size 500) or 1,999 (for samples of size 1,000).  Figure~\ref{fig:normal_pointID} plots empirical rejection probabilities from 1,000 simulation runs.  

From Figure~\ref{fig:normal_pointID} it can be seen that the empirical rejection probabilities are relatively close to the nominal 5\% rejection probability at the boundary of the null region when $\mu_B = 0$.  The behavior of supremum norm tests was identical so only $V_{1n}$ test results are shown.  The $W_{1n}$ and $W_{2n}$ results are close and the differences are due to numerical integration that occurs over one or two dimensions depending on the statistic.

\subsection{Triangular model, identified case}
\begin{figure}[ht] 
  \centering
  \includegraphics[width = 0.9\columnwidth]{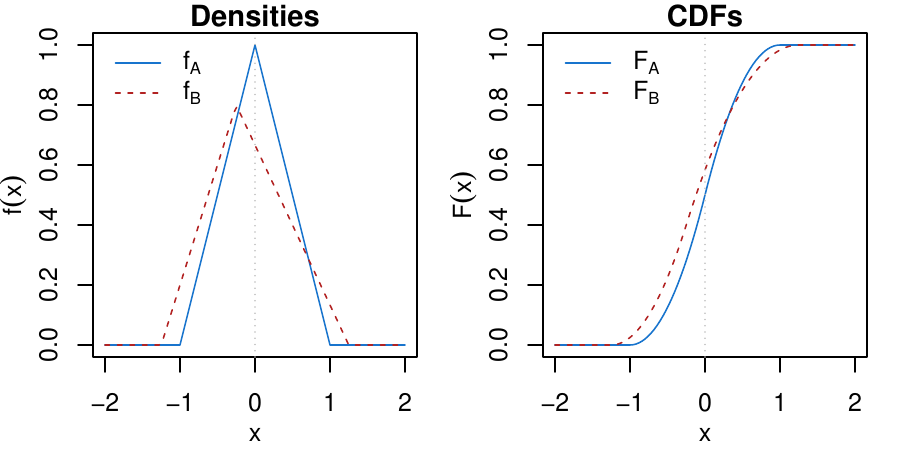}
  \caption{Triangular model densities and distribution functions.  In this example distribution $F_A \succeq_{LASD} F_B$ (in terms of the description in the text, $\epsilon = 1/4$ for distribution $B$).  Heuristically, the higher gains under policy $B$ are outweighed by the probability of larger losses so that distribution $A$ dominates distribution $B$ in the LASD sense, but $F_A \not\succeq_{FOSD} F_B$.} \label{fig:pic_tri}
\end{figure}

In this experiment we use two independent triangular random variables, where we let $\theta = (\alpha, \beta, \gamma)$ denote the lower endpoint of the support, the mode of the distribution and the upper endpoint of the support.  Distribution $A$ uses $\theta_A = (-1, 0, 1)$, while the shape of distribution $B$ is allowed to vary.  For a parameter $\epsilon \in [-1/2, 1/2]$ we let $\theta_B = (-1 - \epsilon / \sqrt{n}, -\epsilon / \sqrt{n}, 1 + \epsilon / \sqrt{n})$, so that all the distributions are local to the boundary of the null region represented by $\epsilon = 0$.  Two distributions are depicted in Figure~\ref{fig:pic_tri}, in which $\epsilon = 1/4$.  This implies that $F_A \succeq_{LASD} F_B$.  From the right panel of the plot it can be seen that these distributions satisfy an LASD ordering, but they would not be ordered by FOSD.

\begin{figure}[ht] 
  \centering
  \includegraphics[width = 0.9\columnwidth]{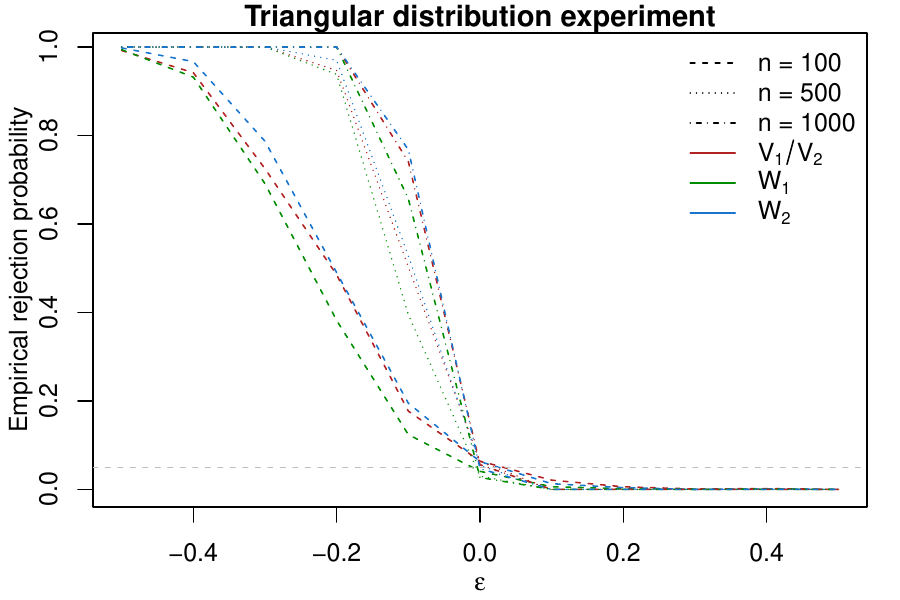}
  \caption{Empirical rejection probabilities of the LASD tests in the point identified triangular model experiment.  The tests are of nominal 5\% size, should have exactly 5\% rejection probability when $\epsilon = 0$ and should reject when $\epsilon < 0$.  Samples of sizes 100, 500 and 1000 correspond respectively to 499, 999 and 1999 bootstrap repetitions.  Distributions are around the boundary of the null region, which is where $\epsilon = 0$, but plotted on an absolute, not local, scale.  1000 simulation repetitions.} \label{fig:tri_pointID}
\end{figure}

Figure~\ref{fig:tri_pointID} shows the empirical rejection results from the triangular model experiment.  We allow $\epsilon$, which controls the shape of distribution $B$, to vary between $-1/2$ and $1/2$.  The tests in this experiment should reject the null when $\epsilon < 0$, should equal the nominal size at $\epsilon = 0$ and should not reject when $\epsilon > 0$.  Because of the restricted supports of the distributions and the relatively small region for $\epsilon$, the horizontal axis for the power curves shown in Figure~\ref{fig:tri_pointID} is the value of the alternative parameters in absolute scale and not local alternatives.  Therefore the power curves show a noticeable change over different values of the sample sizes used.

\subsection{Normal model, partially identified case}
\begin{figure}[ht] 
  \centering
  \includegraphics[width = 0.9\columnwidth]{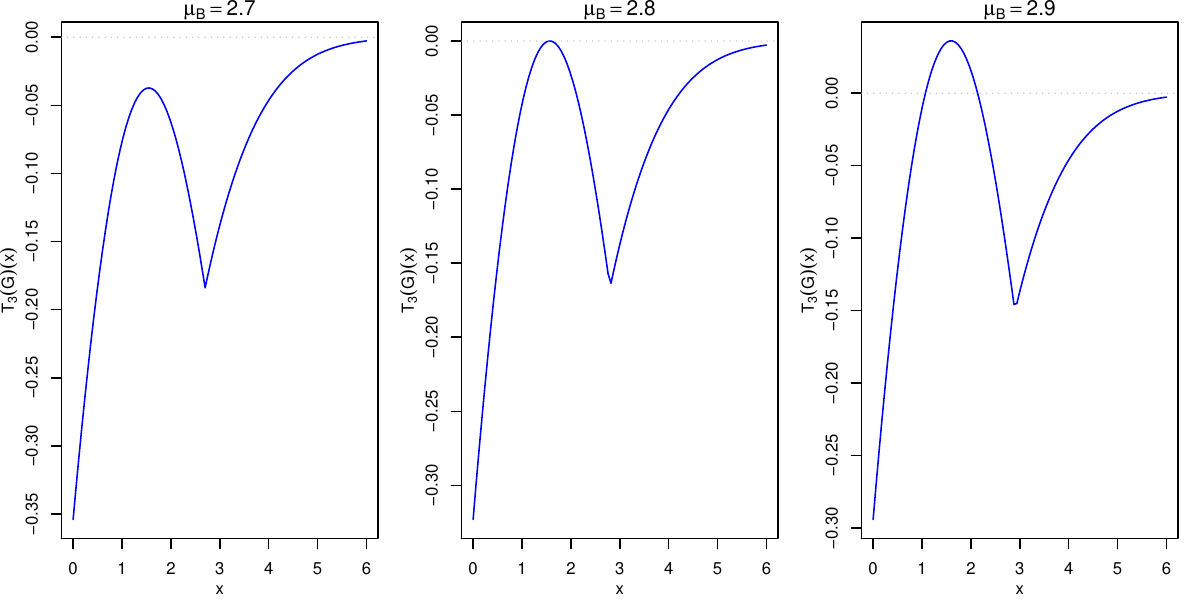}
  \caption{The $T_3(G(\mu_B))$ function for different values of the location of the marginal distribution function $G_B$.  Tests should reject the null hypothesis when $T_3(G)(x) > 0$ for some $x$ as in the right panel.} \label{fig:pic_partial}
\end{figure}

In this experiment we use three independent normal random variables $(Z_0, Z_A, Z_B)$ with scales set to unity and location parameters $\mu = (0, 0, \mu_B)$, where $\mu_B$ is allowed to vary.  We denote this triple of marginal normal CDFs by $G(\mu_B)$.  Rounding to one decimal place, the null $H_0: F_A \succeq_{LASD} F_B$ should be rejected when $\mu_B > 2.8$.  We let $\mu_B$ vary locally around this approximate boundary point.  Figure~\ref{fig:pic_partial} depicts the $T_3(G(\mu_B))$ function for $\mu_B = 2.7, 2.8$ or $2.9$.  Tests are designed to detect the positive deviation in the right-most panel of the figure, when $T_3(G)(x) > 0$ for some $x \geq 0$.

\begin{figure}[ht] 
  \centering
  \includegraphics[width = 0.9\columnwidth]{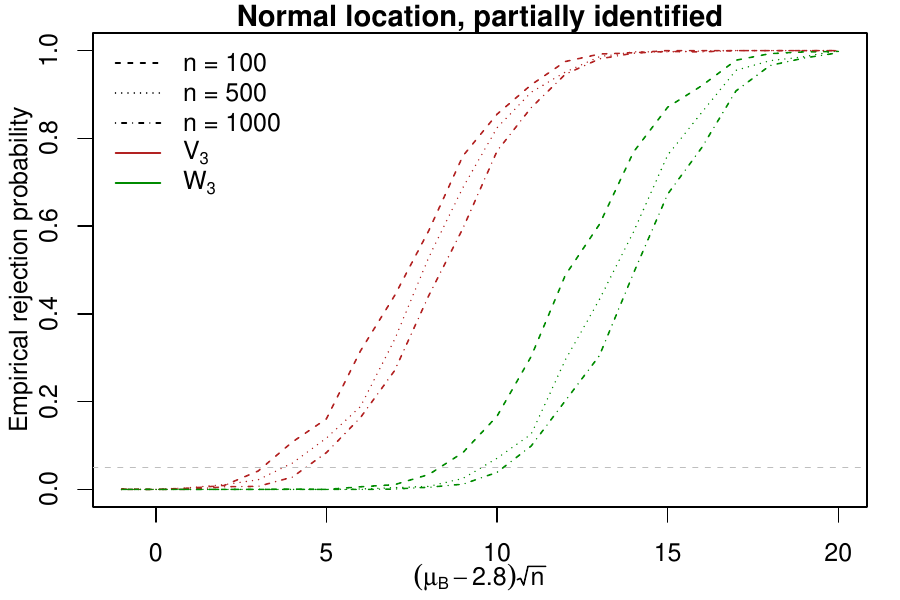}
  \caption{Empirical rejection probabilities of the LASD tests in the partially identified normal location model experiment.  The control and policy $A$ distributions have means set to zero, while the location of policy $B$ is allowed to vary.  The tests are of nominal 5\% size, should have exactly 5\% rejection probability when $(\mu_B - 2.8) \sqrt{n} = 0$ and should reject when $(\mu_B - 2.8) \sqrt{n} > 0$ (alternatives are local to the boundary of the set $\calP_0^{nec}$ described in the text).  Samples of sizes 100, 500 and 1000 correspond respectively to 499, 999 and 1999 bootstrap repetitions.  1000 simulation repetitions.} \label{fig:partialID_normal}
\end{figure}

Figure~\ref{fig:partialID_normal} shows empirical rejection probabilities for tests with three independent normal distributions.  The tests are not conducted under any assumptions about the independence of the samples.  The rejection probabilities are different than those in the point-identified experiments~--- more evidence is needed to detect deviations from the null region than in the identified case, because the bound $\mathrm{U}_B$ combines observations from the control and sample $B$.  Although more information is necessary, it is important to note that these alternatives (like in the other experiments) are local to the boundary of the $\calP_0^{nec}$ set.

As can be seen in Figure~\ref{fig:partialID_normal}, the tests in the partially identified case do not reject the null with as high a probability as in the point identified case, which is a direct result of the lack of knowledge about inter-sample correlations that dictates the form of the $T_3$ function defined in the main text.  Also, it appears as though these deviations from the null are not very well detected by the Cram\'er-von Mises tests in relation to the Kolmogorov-Smirnov tests.  However, it is important to note that in this example, alternatives are local alternatives, and represent smaller and smaller deviations from the null region as sample sizes increase.

\section{Application}

In this section we present the additional test results for the empirical application discussed in Section 5 of the main paper.  We show $V_{2n}$ test statistics and reproduce the table of $W_{2n}$ results used in the main text so that they may be compared.  The tests have very similar qualitative conclusions.  Tests based on $V_{1n}$ and $W_{1n}$ are not shown because they are identical or nearly identical to tests based on $V_{2n}$ and $W_{2n}$.

\begin{small}
\input{"supp_extras/all_tests.tex"}
\end{small}

Finally, we note that the example could be used to conduct tests under partial identification, as if we had no knowledge of the longitudinal structure of the data.  However, tests using $V_{3n}$ or $W_{3n}$ statistics were all identically zero and had p-values equal to 1, and the table of corresponding results is omitted.

\subsection{Inferring dominance from partially-identified treatment distributions} \label{sec:partialID}

In this section we extend dominance tests to the case that distribution functions $F_A$ and $F_B$ are only partially identified by their Makarov bounds.  Suppose that $Z_0$, $Z_A$ and $Z_B$ are random variables with marginal distribution functions $G = (G_0, G_A, G_B)$, but the joint probability distribution $P$ of the vector $(Z_0, Z_A, Z_B)$ is unknown, so that $F_A$ and $F_B$ are not point identified because they are the unknown distribution functions of $X_A = Z_A - Z_0$ and $X_B = Z_B - Z_0$.  Nevertheless, we wish to test the hypotheses in~\eqref{hypotheses_general}, which depend on $F_A$ and $F_B$.

\subsubsection{Test statistics}

Recall equations~\eqref{ptfsdsuff} and~\eqref{ptfsdnec} from Section~\ref{sec:ptsd}.  Restated in terms of the null hypothesis $F_A \succeq_{LASD} F_B$, condition \eqref{ptfsdsuff} is sufficient to imply the null hypothesis is true, while \eqref{ptfsdnec} represents a necessary condition for dominance.  Denote by $\calP^{suf}$ the set of distributions that satisfy \eqref{ptfsdsuff} and let $\calP^{nec}$ collect all distributions that satisfy \eqref{ptfsdnec}.  Then still using the label $\calP_0$ for the set of distributions such that $X_A$ dominates $X_B$, we have the (strict) inclusions $\calP^{suf} \subset \calP_0 \subset \calP^{nec}$.  Given this relation, without any further identification conditions, we look for significant violations of the necessary condition, since $P \notin \calP^{nec}$ implies $P \notin \calP_0$. This generally results in trivial power against $P \in \calP^{nec} \backslash \calP_0$, but it avoids overrejection.

To test the null \eqref{hypotheses_general} we employ the inequality specified in equation \eqref{ptfsdnec} from Theorem~\ref{thm:boundsfsd}.  For each $x \in \calX$ let
\begin{equation}\label{t3def}
  T_3(G)(x) = L_A(-x) + L_A(x) - U_B(-x) - U_B(x),
\end{equation}
where $L_A$ and $U_B$ are defined in~\eqref{lowerk} and~\eqref{upperk}.  To see the explicit dependence of $T_3$ on $G$, rewrite~\eqref{t3def}, using the identity $\inf f = - \sup (-f)$ in the definition of $U_B$ as
\begin{multline} \label{t3_detail}
  T_3(G)(x) = \sup_{u \in \R} (G_A(u) - G_0(u + x)) + \sup_{u \in \R} (G_A(u) - G_0(u - x)) \\
  - 2 + \sup_{u \in \R} (G_0(u + x) - G_B(u)) + \sup_{u \in \R} (G_0(u - x) - G_B(u)).
\end{multline}
As before, $T_3$ has been written in such a way that a violation of the null hypothesis $F_A \succeq_{LASD} F_B$ is indicated by observing some $x$ such that $T_3(G)(x) > 0$.

The above map shares a similar feature with the $T_1$ map in the previous section~--- the marginal (in $u$) optimization maps are pointwise directionally differentiable at each $x \geq 0$, but $f(u, x) \mapsto \sup_u f(u, x)$ is not Hadamard differentiable as a map from $\ell^\infty(\R \times \calX)$ to $\ell^\infty(\calX)$ due to lack of uniform convergence to the pointwise derivatives.  One solution to this problem is the same as it was when considering statistics based on the plug-in estimate of $T_1(F)$: examine the distribution of test functionals applied to the process, which are Hadamard directionally differentiable (shown in Lemma~\ref{lem:sup_add} in Appendix~\ref{app:hadamard}).  

Given observed samples $\{Z_{ki}\}$ for $k \in \{0, A, B\}$, define the marginal empirical distribution functions $\bbG_n = (\bbG_{0n}, \bbG_{An}, \bbG_{Bn})$, where $\bbG_{kn}(z) = \frac{1}{n_k} \sum_i \one\{Z_{ki} \leq z\}$ for $k \in \{0, A, B\}$, and let $\bbL_{An}$ and $\bbU_{Bn}$ be the plug-in estimates of the bounds: for each $x \in \calX$, let
\begin{align*}
  \bbL_{An}(x) &= L(x, \bbG_{0n}, \bbG_{An}) \\
  \bbU_{Bn}(x) &= U(x, \bbG_{0n}, \bbG_{Bn}),
\end{align*}
where the maps $L$ and $U$ were introduced in equations \eqref{lowerk} and \eqref{upperk}.  To estimate $T_3$ in \eqref{t3def} we use the plug-in estimate $T_3(\bbG_n)$.  As in the previous section, we consider the following Kolmogorov-Smirnov and Cram\'er-von Mises type test statistics:
\begin{align}
  V_{3n} &= \sqrt{n} \sup_{x \in \calX} (T_3(\bbG_n)(x))^+ \label{V3n_def} \\
  W_{3n} &= \sqrt{n} \left( \int_\calX \left( (T_3(\bbG_n)(x))^+ \right)^2 \dd x \right)^{1/2}. \label{W3n_def}
\end{align}
The next subsections establish limiting distributions for $V_{3n}$ and $W_{3n}$ and suggest a resampling procedure to estimate the distributions.

\subsubsection{Limiting distributions}
Once again, it is necessary to define the region where the test statistics have nontrivial distributions.  We say that distribution $P \in \calP_{00}^{nec}$ when $\sup_{x \in \text{int} \calX} T_3(G)(x) = 0$.  As mentioned at the beginning of the section, $\calP_{00}^{nec}$ is not the set of $P$ such that $F_A \succeq_{LASD} F_B$, rather those that satisfy this necessary condition, or in other words, $\calP_0 \subset \calP^{nec}$.  There is no obvious connection between $\calP_0$ and $\calP_{00}^{nec}$~--- the $P$ in $\calP_{00}^{nec}$ are simply those that lead to nontrivial asymptotic behavior of the $T_3$ statistic, as will be shown in Theorem~\ref{thm:partial_teststats}.   Define the contact set for the $T_3$ criterion function by
\begin{equation*}
  \calX_0^{nec}(P) = \left\{ x \in \text{int} \calX : T_3(G)(x) = 0 \right\}.
\end{equation*}
Next, we define a few functions that are analogous to the $m_1$ and $m_2$ used in the point-identified case, and which come from separating equation~\eqref{t3_detail} into four sub-functions.  Let $m_1(u, x) = G_A(u) - G_0(u + x)$, $m_2(u, x) = G_A(u) - G_0(u - x)$, $m_3(u, x) = G_0(u + x) - G_B(u)$ and $m_4(u, x) = G_0(u - x) - G_B(u)$.  These functions are used to define, for $k = 1, \ldots 4$, for any $x \in \calX$ and $\epsilon \geq 0$, the set-valued maps
  \begin{equation} \label{epsmax_sets}
    \calM^k(x, \epsilon) = \left\{ u \in \R : m_k(u, x) \geq \sup_{u \in \R} m_k(u, x) - \epsilon \right\}.
  \end{equation}
  Also for the supremum norm statistic another relevant set of $\epsilon$-maximizers exists: for any $\epsilon \geq 0$, let
  \begin{equation}
  \calM^{nec}(\epsilon) = \left\{ (u, x) \in \R \times \calX : \sum_{k=1}^4 m_k(u, x) \geq \sup_{u, x} \sum_{k=1}^4 m_k(u, x) - \epsilon \right\}.
  \end{equation}
  Under the null hypothesis that the supremum is zero, $\lim_{\epsilon \searrow 0} \calM^{nec}(\epsilon) = \calX_0^{nec}$ in the Painlev\'e-Kuratowski sense, as seen in the expression for $V_3$ in the next theorem.

Now we turn to regularity assumptions on the observed data.  The only difference between these assumptions and assumptions~\ref{assumptionA_first}-\ref{assumptionA_last} is that we must now make assumptions for three samples instead of two.
\begin{enumerate}[label=\textbf{B\arabic*}]
  \item \label{assumptionB_first} The observations $\{Z_{0i}\}_{i=1}^{n_0}$, $\{Z_{Ai}\}_{i=1}^{n_A}$ and $\{Z_{Bi}\}_{i=1}^{n_B}$ are iid samples and independent of each other and are continuously distributed with marginal distribution functions $G_0$, $G_A$ and $G_B$ respectively.
  \item \label{assumptionB_last} The sample sizes $n_0$, $n_A$ and $n_B$ increase in such a way that $n_k / (n_0 + n_A + n_B) \rightarrow \lambda_k$ as $n_0, n_A, n_B \rightarrow \infty$, for $k \in \{0, A, B\}$, where $0 < \lambda_k < 1$.  Let $n = n_0 + n_A + n_B$.
  %\item \label{assumptionB_last} The evaluation set $\calX \subseteq \mathrm{supp}(G_0) \cup \mathrm{supp}(G_A) \cup \mathrm{supp}(G_B)$, the supports of the marginal distribution functions.
\end{enumerate}

Before stating the next theorem, it is convenient to make some definitions.  Under assumptions~\ref{assumptionB_first}-\ref{assumptionB_last}, standard results in empirical process theory show that there is a Gaussian process $\calG_G$ such that $\sqrt{n}(\bbG_n - G) \cw \calG_G$ \citep[Example 19.6]{vanderVaart98}.  For each $(u, x)$, denote the transformed empirical processes and their (Gaussian) limits
  \begin{equation} \label{transformed_limits}
    \begin{aligned}
    \sqrt{n}(\bbG_{An}(u) - \bbG_{0n}(u + x) - G_A(u) + G_0(u + x)) = \bbG_{1n}(u, x) \cw \calG_1(u, x) \\ 
    \sqrt{n}(\bbG_{An}(u) - \bbG_{0n}(u - x) - G_A(u) + G_0(u - x)) = \bbG_{2n}(u, x) \cw \calG_2(u, x) \\
    \sqrt{n}(\bbG_{0n}(u + x) - \bbG_{Bn}(u) - G_0(u + x) + G_B(u)) = \bbG_{3n}(u, x) \cw \calG_3(u, x) \\
    \sqrt{n}(\bbG_{0n}(u - x) - \bbG_{Bn}(u) - G_0(u - x) + G_B(u)) = \bbG_{4n}(u, x) \cw \calG_4(u, x)
    \end{aligned}
  \end{equation}

Given the above and definitions, the asymptotic behavior of $V_{3n}$ and $W_{3n}$ can be established.  
\begin{theorem} \label{thm:partial_teststats}
  Under assumptions~\ref{assumptionB_first}-\ref{assumptionB_last}:
  \begin{enumerate}
    \item Suppose that $P \in \calP_{00}^{nec}$.  As $n \rightarrow \infty$, $V_{3n} \cw V_3$ and $W_{3n} \cw W_3$, where, given the definitions~\eqref{transformed_limits} and~\eqref{epsmax_sets},
      \begin{equation*}
        V_3 = \left( \sup_{x \in \calX_0^{nec}(P)} \sum_{k=1}^4 \lim_{\epsilon \searrow 0} \sup_{u \in \calM^k(x, \epsilon)} \calG_k(u, x) \right)^+
      \end{equation*}
      and
      \begin{equation*}
        W_3 = \left( \int_{\calX_0^{nec}(P)} \left( \left( \sum_{k=1}^4 \lim_{\epsilon \searrow 0} \sup_{u \in \calM^k(x, \epsilon)} \calG_k(u, x) \right)^+ \right)^2 \dd x \right)^{1/2}.
      \end{equation*}
    \item Suppose that $P \in \calP^{nec} \backslash \calP_{00}^{nec}$.  Then as $n \rightarrow \infty$, $\prob{V_3 > \epsilon} \rightarrow 0$ and $\prob{W_3 > \epsilon} \rightarrow 0$ for all $\epsilon > 0$.
    \item Suppose that $P \notin \calP^{nec}$.  Then as $n \rightarrow \infty$, $\prob{V_3 > c} \rightarrow 1$ and $\prob{W_3 > c} \rightarrow 1$ for all $c \geq 0$.
  \end{enumerate}
\end{theorem}
The results of this theorem parallel those in Theorem~\ref{thm:obs_teststats}.  The distributions of these test statistics are complex.  Unfortunately, the limits of the $\epsilon$-maximization operations in the first part of the theorem cannot be simplified because they are part of the definition of the CDF bound functions, and the fact that the sum of four suprema is nonpositive does not imply that each supremum is nonpositive.  A consistent resampling procedure for inference is discussed in the next subsection.  The conservatism of these tests is reflected in the second part above.  There may be $P \notin \calP_0$ such that $P \in \calP^{nec} \backslash \calP_{00}^{nec}$, meaning the test will not detect that this distribution violates the hypothesis that $F_A \succeq_{LASD} F_B$.  

\subsubsection{Resampling procedures for inference under partial identification}
Now we turn to the issue of conducting practical inference using estimated bound functions and the necessary condition for LASD.  As before, resampling can be implemented by estimating the derivatives of either $V_3$ or $W_3$.  These estimates represent the only major difference from the resampling scheme developed in the point identified setting.

The estimates required for tests based on $V_{3n}$ and $W_{3n}$ are similar to those used in the point-identified case.  Define a grid of values $\mathbb{X} \subset \R$ and let $\mathbb{X}^+$ be the sub-grid of nonnegative points such that $\mathbb{X}^+ \subset \calX$.  We suggest a grid because otherwise these functions may need to be evaluated over a prohibitive number of points because each upper and lower bound function may take unique values at all pairs of sample observations (each bound function is constructed using two samples).  The size of this grid should be as large as can be tolerated in order to approach the supremum over $\calX$.  In contrast, for point-identified tests with plug-in empirical CDFs it is sufficient to evaluate the criterion functions at the union of the two sample observations. For a sequence $a_n$ such that $a_n \searrow 0$ and $\sqrt{n}a_n \rightarrow \infty$, define the estimate of the contact set
\begin{equation} \label{est_contact_set_nec}
  \hat{\calX}_0^{nec} = \left\{ x \in \mathbb{X}^+ : |\bbL_{An}(-x) + \bbL_{An}(x) - \bbU_{Bn}(-x) - \bbU_{Bn}(x)| \leq a_n \right\}.
\end{equation}
When this estimated set is empty, set $\hat{\calX}_0^{nec} = \mathbb{X}^+$.  The inner maximization step that occurs in the definition of the test statistics requires an estimate of the $\epsilon$-maximizers of each sub-process, that is, estimates of~\eqref{epsmax_sets} for $k = 1, \ldots 4$.  For these sets we also use the same sort of estimator: for $\{b_n\}$ such that $b_n \searrow 0$ and $\sqrt{n}b_n \rightarrow \infty$, for each $x \in \mathbb{X}^+$ let
\begin{equation} \label{epsmaxA_est}
  \hat{\calM}^k(x) = \left\{ u \in \mathbb{X} : \hat{m}_{kn}(u, x) \geq \max_{u \in \mathbb{X}} \hat{m}_{kn}(u, x) - b_n \right\}
\end{equation}
where the $\hat{m}_{kn}$ are plug-in estimators of $m_k$.  Finally, for a sequence $d_n$ such that $d_n \searrow 0$ and $\sqrt{n}d_n \rightarrow \infty$, define the estimator
  \begin{equation} \label{epsmaxB_est}
  \hat{\calM}^{nec} = \left\{ (u, x) \in \mathbb{X} \times \mathbb{X}^+ : \sum_{k=1}^4 \hat{m}_{kn}(u, x) \geq \max_{(u, x) \in \mathbb{X} \times \mathbb{X}^+} \sum_{k=1}^4 \hat{m}_{kn}(u, x) - d_n \right\}.
  \end{equation}
Putting these estimates together, we find the derivative estimates described in the resampling scheme below.

\noindent \textbf{Resampling routine to estimate the distributions of $V_{3n}$ and $W_{3n}$}
\begin{enumerate}
  \item \label{tcontact3} If using a Cram\'er-von Mises statistic, given a sequence of constants $\{a_n\}$, estimate the contact set $\hat{\calX}_0^{nec}$.  If using a Kolmogorov-Smirnov statistic, given sequences of constants $\{b_n\}$ and $\{d_n\}$, estimate $\hat{\calM}^k(\cdot)$ for $k = 1, \ldots 4$ and $\hat{\calM}^{nec}$.
\end{enumerate}
Next repeat the following two steps for $r = 1, \ldots, R$:
\begin{enumerate}
  \setcounter{enumi}{2}
  \item Construct the resampled processes $\calG_{kn}^* = \sqrt{n}(\bbG_{kn}^* - \bbG_{kn})$ using an exchangeable bootstrap.
  \item Calculate the resampled test statistic
    \begin{equation*}
      V_{r3n}^* = \left( \max_{x \in \hat{\calM}^{nec}} \sum_{k=1}^4 \max_{u \in \hat{\calM}^k(x)} \calG_{kn}^*(u, x) \right)^+
    \end{equation*}
    or
    \begin{equation*}
      W_{r3n}^* = \left( \int_{\hat{\calX}_0^{nec}} \left( \left( \sum_{k=1}^4 \max_{u \in \hat{\calM}^k(x)} \calG_{kn}^*(u, x) \right)^+ \right)^2 \dd x \right)^{1/2}.
    \end{equation*}
\end{enumerate}
Finally,
\begin{enumerate}
  \setcounter{enumi}{5}
  \item Let $\hat{q}_{V^{*}_{3}}(1-\alpha)$ and $\hat{q}_{W^{*}_{3}}(1-\alpha)$ be the $(1-\alpha)^{\text{th}}$ sample quantile from the bootstrap distributions of $\{V_{r3n}^*\}_{r=1}^R$ or $\{W_{r3n}^*\}_{r=1}^R$, respectively, where $\alpha \in (0, 1)$ is the nominal size of the tests. We reject the null hypothesis \eqref{hypotheses_T1} if $V_{3n}$ and $W_{3n}$ defined in~\eqref{V3n_def} or~\eqref{W3n_def} are, respectively, larger than $\hat{q}_{V^{*}_{3}}(1-\alpha)$ or $\hat{q}_{W^{*}_{3}}(1-\alpha)$.
\end{enumerate}

The following theorem guarantees that the resampling scheme is consistent on $\calP_{00}^{nec}$.
\begin{theorem} \label{thm:resample_consistent_intervalID}
  Make assumptions~\ref{assumptionB_first}-\ref{assumptionB_last} and suppose that $P \in \calP^{nec}_{00}$. %Let $\hat{q}_{V^*_3}(1-\alpha)$ and $\hat{q}_{W^*_3}(1-\alpha)$ be the $(1-\alpha)^{\text{th}}$ sample quantile from the bootstrap distributions as described in the routines above.
  Let $X$ denote the sample observations.  Then the bootstrap is consistent: 
  \begin{equation*}
    \sup_{f \in BL_1} \left| \ex{f(V_{3n}^*) | X} - \ex{f(V_3)} \right| = o_P(1)
  \end{equation*}
  and
  \begin{equation*}
    \sup_{f \in BL_1} \left| \ex{f(W_{3n}^*) | X} - \ex{f(W_3)} \right| = o_P(1).
  \end{equation*}
  In particular, when $P \in \calP_{00}^{nec}$ the resampling procedure outlined above results in asymptotically valid inference: for any $P \in \calP_{00}^{nec}$, letting $q_{V^*_3}(1 - \alpha) = \lim_{R \rightarrow \infty} \hat{q}_{V^*_3}(1 - \alpha)$ and $q_{W^*_3}(1 - \alpha) = \lim_{R \rightarrow \infty} \hat{q}_{W^*_3}(1 - \alpha)$,
  \begin{equation*}
    \limsup_{n \rightarrow \infty} \prob{ V_{3n} > q_{V_3^*}(1 - \alpha) } \leq \alpha
  \end{equation*}
  and
  \begin{equation*}
    \limsup_{n \rightarrow \infty} \prob{ W_{3n} > q_{W_3^*}(1 - \alpha) } \leq \alpha.
  \end{equation*}
\end{theorem}

Like in the point-identified setting, we define a resampling scheme and state Theorem~\ref{thm:resample_size_intervalID} under the imposition of the hypothesis that $P \in \calP^{nec}_{00}$.  The testing procedure based on the $T_3$ criterion function controls size over all $P \in \calP^{nec}$, a superset of $\calP_0$.  The size of the resampling inference scheme for $P \in \calP^{nec}$ and local alternatives is stated formally in Theorem~\ref{thm:resample_size_intervalID} in Appendix~\ref{app:hadamard}.  However, using only a necessary condition for inference comes at a cost, which is the possibility of trivial power against some alternative $P \notin \calP_0$.  For any $P \in \calP^{nec} \backslash \calP_0$, the probability of rejecting the null is also less than or equal to $\alpha$.  More generally, results about size and power against various alternatives that can be specified for point identified distributions are not available for the partially identified case.  On the other hand, the test controls size on $\calP_0$, which is a set of treatment outcome distributions that cannot be observed directly.

Now we consider local size control using the resampling routine outlined above to test the null hypothesis that $F_A \succeq_{LASD} F_B$ when the distributions are only partially identified.  It is no longer possible to guarantee exact rejection probabilities because the test is based on a superset of $\calP_0$, but we can still show that the test does not overreject.
\begin{theorem} \label{thm:resample_size_intervalID}
  Make assumptions~\ref{assumptionB_first}-\ref{assumptionB_last}.  Also assume that $\calX$ is a convex set.  Let $\hat{q}_{V^*_3}(1-\alpha)$ and $\hat{q}_{W^*_3}(1-\alpha)$ be the $(1-\alpha)^{\text{th}}$ sample quantile from the bootstrap distributions of $\{V_{r3n}^*\}_{r=1}^R$ or $\{W_{r3n}^*\}_{r=1}^R$ as described in the routine above, and let $q_{V^*_3}(1-\alpha) = \lim_{R \rightarrow \infty} \hat{q}_{V^*_3}(1 - \alpha)$ and $q_{W^*_3}(1-\alpha) = \lim_{R \rightarrow \infty} \hat{q}_{W^*_3}(1 - \alpha)$.  When the sequence of alternative distributions $P_n$ satisfy~\eqref{local_alt_def} and $T_3(F_n)(x) \leq 0$ for all $x \geq 0$,
  \begin{equation*}
    \limsup_{n \rightarrow \infty} P_n \left\{ V_{3n} > q_{V^*_3}(1-\alpha) \right\} \leq \alpha 
  \end{equation*}
  and 
  \begin{equation*}
    \limsup_{n \rightarrow \infty} P_n \left\{ W_{3n} > q_{W^*_3}(1-\alpha) \right\} \leq \alpha.
  \end{equation*}
\end{theorem}

\begin{proof}[Proof of Theorem~\ref{thm:partial_teststats}]
Consider $V_3$ first.  Note that $V_{3n}$ can be rewritten as 
\begin{equation*}
V_{3n} = \sqrt{n}\sup (T_3(\bbG_n))^+ = \sqrt{n} \max\{0, \sup T_3(\bbG_n)\}.
\end{equation*}
Lemma~\ref{lem:sup_add}, extended to the four parts of the $T_3$ process, and the condition that $\calX_0^{nec}(P) \neq \ns$, implies each of the four inner results.  The derivative of the positive-part map discussed in~\eqref{phiprime_def}, with the hypothesis that $P \in \calP_{00}^{nec}$, which implies $\lim_{\epsilon \searrow 0} \calM^{nec}(\epsilon) = \calX_0^{nec}$ (in the Painlev\'e-Kuratowski sense, similar to the convergence discussed in the proof of Theorem~\ref{thm:obs_teststats}), and the chain rule imply the outer part of the derivative and Theorem 2.1 of \citet{FangSantos19} implies the result.  For $W_{3n}$ and $W_3$, the finite-sample integrand converges pointwise for each $x \in \calX$ to the limit.  By assumption there are no $x$ such that the integrand is positive, which leaves the $x$ in $\calX_0^{nec}(P)$ as the nontrivial part of the integral.  Because the limit is assumed square-integrable, dominated convergence, Lemma~\ref{lem:lpnorm_hd} and Theorem 2.1 of \citet{FangSantos19} imply the result.

  For Part 2, note that by hypothesis $\calX_0^{nec}(P) = \varnothing$ and there are no $x$ that result in $T_3(G)(x) > 0$.  Therefore Theorem 2.1 of \citet{FangSantos19}, along with the chain rule, Lemmas~\ref{lem:sup_add} and~\ref{lem:lpnorm_hd} and the positive-part map, imply the result.  The proof of Part 3 is the same as the analogous part of the proof of Theorem~\ref{thm:obs_teststats}.
\end{proof}

\begin{proof}[Proof of Theorem~\ref{thm:resample_consistent_intervalID}]
  For both statistics, Assumptions 1-3 of \citet{FangSantos19} are trivially satisfied \citep[Theorem 2.8.4]{vanderVaartWellner96} or satisfied by construction in the case of the bootstrap weights.  Below we check that their Assumption 4 is also satisfied for both statistics, so that the statement of the theorem follows from their Theorem 3.2.

  Consider $V_{3n}$ first, and write the supremum statistic as a function of underlying processes abstractly labeled $g$: the limiting variable relies (through the delta method) on a map of the form $V_3 = V_3(g) = (\phi'_{\theta(g)} \circ \theta'_g) (h)$, where $g \in (\ell^\infty(\R \times \calX))^4$, $\phi'_x$ is defined in~\eqref{phiprime_def} and $\theta'_g$ in~\eqref{thetaprime_def} (extended to four functions as the arguments of the map).  $V_{3n}$ uses the sample estimates of these functions.  Under the null hypothesis $\theta(g) = 0$, so that we may estimate $\hat{\phi}_n'(x) = (x)^+$, which is Lipschitz because $|(x)^+ - (y)^+| \leq |x - y|$.  Writing the formula for the estimate of the derivative of $\theta$ for just two functions $f$ and $g$ (since the estimator for four functions can be extended immediately from this case), we have, given sequences $\{b_n\}$ and $\{d_n\}$,
  \begin{equation*}
    \hat{\theta}'(h, k) = \max_{x \in \hat{\calM}_\theta} \left( \max_{u \in \hat{\calM}_f(x)} h(u, x) + \max_{u \in \hat{\calM}_g(x)} k(u, x) \right).
  \end{equation*}
  This map is Lipschitz in $(h, k)$: given any $(f,g)$ pair, paraphrasing the sets over which maxima are taken and their arguments, we have
  \begin{align*}
    \left| \hat{\theta}'(h_1, k_1) - \hat{\theta}'(h_2, k_2) \right| &= \left| \max_{\hat{\calM}_\theta} \left( \max_{\hat{\calM}_f} h_1 + \max_{\hat{\calM}_g} k_1 \right) - \max_{\hat{B}} \left( \max_{\hat{\calM}_f} h_1 + \max_{\hat{\calM}_g} k_1 \right) \right| \\
    {} &\leq \max_{\hat{\calM}_\theta} \left| \max_{\hat{\calM}_f} h_1 + \max_{\hat{\calM}_g} k_1 - \max_{\hat{\calM}_f} h_2 - \max_{\hat{\calM}_g} k_2 \right| \\
    {} &\leq \max_{\hat{\calM}_\theta} \max_{\hat{\calM}_f} |h_1 - h_2| + \max_{\hat{\calM}_\theta} \max_{\hat{\calM}_g} |k_1 - k_2| \\
    {} &\leq 2 \max\left\{ \|h_1 - h_2\|_\infty, \|k_1 - k_2\|_\infty \right\} \\
    {} &= 2 \| (h_1, k_1) - (h_2, k_2) \|_\infty.
  \end{align*}
  Because all the maps in the chain that defines $V_{3n}$ are Lipschitz, $V_{3n}$ is itself Lipschitz, and therefore Lemma S.3.6 of \citet{FangSantos19} implies that their Assumption 4 holds if $\|(\hat{\phi}'_{\theta(g)} \circ \hat{\theta}'_g) (h) - (\phi'_{\theta(g)} \circ \theta'_g) (h) \| = o_P(1)$ (where the arguments $g$ and $h$ are again elements of $(\ell^\infty(\R \times \calX))^4$).  This follows from the consistency of the $\epsilon$-maximizer estimates.

  Next consider $W_{3n}$.  For this part simplify the inner part to the sum of two functions, $f$ and $g$, since the result is a simple generalization.  Write $W_{3n} = W_{3n}(h, k) = (\hat{\lambda}'_{\mu(f, g)} \circ \hat{\mu}'_{f,g})(h, k)$, where the marginal (in $u$) maximization map $\mu$ is defined for each $x \geq 0$, by $\mu(f, g)(x) = \sup_\calU f(u, x) + \sup_\calU g(u, x)$ and for each $x \geq 0$, $\hat{\mu}_{f,g}'(h,k)(x) = \max_{u \in \hat{\calM}_f(x)} h(u, x) + \max_{u \in \hat{\calM}_g(x)} k(u, x)$ (define $\hat{\calM}_f(x)$ and $\hat{\calM}_g(x)$ as in~\eqref{epsmaxA_est}).  First,
  \begin{align*}
    \left\| \hat{\mu}'(h_1, k_1) - \hat{\mu}'(h_2, k_2) \right\|_\infty &= \sup_\calX \left| \max_{\hat{\calM}_f(x)} h_1 + \max_{\hat{\calM}_g(x)} k_1 - \max_{\hat{\calM}_f(x)} h_2 - \max_{\hat{\calM}_g(x)} k_2 \right| \\
    {} &\leq \|h_1 - h_2\|_\infty + \|k_1 - k_2\|_\infty \\
    {} &\leq 2\|(h_1, k_1) - (h_2, k_2)\|_\infty.
  \end{align*}
  Second, for square integrable $f$ and $h$ consider the estimate, assuming $P \in \calP^{nec}$,
  \begin{equation*}
    \hat{\lambda}'(h) = \lambda(h|_{\hat{\calX}_0})
  \end{equation*}
  where $f|_A$ denotes the restriction of the function $f$ to the set $A$.  On $\hat{\calX}_0$ the subadditivity of the norm trivially implies that $\hat{\lambda}'$ is Lipschitz there.  %On $\hat{\calX}_+$, we have
%  \begin{align*}
%    \frac{1}{\lambda(f|_{\hat{\calX}_+})} \int_{\hat{\calX}_+} f(x) (h_1(x) - h_2(x)) \dd x &\leq \frac{\lambda(f|_{\hat{\calX}_+})}{\lambda(f|_{\hat{\calX}_+})} \|h_1 - h_2\|_\infty \\
%    {} &= \|h_1 - h_2\|_\infty.
%  \end{align*}
This implies that $\hat{\lambda}'$ is a Lipschitz map, and in turn that $\hat{\lambda}'_{\mu(f, g)} \circ \hat{\mu}'_{f,g}$ is Lipschitz.

  Finally, $\hat{\mu}_{f,g}'(h,k)(x)$ converges for each $x$ the pointwise limit 
  \begin{equation*}
    \mu_{f,g}'(h,k)(x) = \lim_{\epsilon \searrow 0} \left( \sup_{u \in \hat{\calM}_f(x, \epsilon)} h(u, x) + \max_{u \in \hat{\calM}_g(x, \epsilon)} k(u, x) \right).
  \end{equation*}
  The set estimators $\hat{\calX}_0$ and $\hat{\calX}_+$ are consistent estimators for $\calX_0$ and $\calX_+$ using the same argument as above for the supremum norm.  Then for square integrable $h$ and $k$, the dominated convergence theorem implies that for any given $f,g$, 
  \begin{equation*}
    \left| (\hat{\lambda}'_{\mu(f, g)} \circ \hat{\mu}'_{f,g})(h, k) - (\lambda'_{\mu(f, g)} \circ \mu'_{f,g})(h, k) \right| = o_P(1),
  \end{equation*}
  and Lemma S.3.6 of \citet{FangSantos19} implies the result.  The last part is implied by Theorem~\ref{thm:resample_size_intervalID}.
\end{proof}

\section{Results on differentiability and size control} \label{app:hadamard}
This section includes a definition and short discussion of the Hadamard directional differentiability concept and contains important intermediate results on Hadamard derivatives used to establish the main results in the text.  Next we present some results on the control of size over the null region using the proposed resampling methods.  Finally, there is one remark regarding the computation of $T_1$ and $T_2$ processes ($T_3$ processes should probably be computed on a grid for the sake of computation time).  Proof of the results discussed in this appendix are collected in Appendix~\ref{app:appA_proof}.

\subsection{Hadamard differentiability and statistical inference}
The Hadamard derivative is a standard tool used to analyze the asymptotic behavior of nonlinear maps in empirical process theory \citep[Section 20.2]{vanderVaart98}.  We provide a definition here for completeness, along with its directional counterpart.
\begin{deff}[Hadamard differentiability]\label{def:diff}
  Let $\bbD$ and $\bbE$ be Banach spaces and consider a map $\phi: \bbD_\phi \subseteq \bbD \rightarrow \bbE$.
  \begin{enumerate}
    \item $\phi$ is \emph{Hadamard differentiable} at $f \in \bbD_\phi$ tangentially to a set $\bbD_0 \subseteq \bbD$ if there is a continuous linear map $\phi': \bbD_0 \rightarrow \bbE$ such that 
      \begin{equation*}
        \lim_{n \rightarrow \infty} \left\| \frac{\phi(f + t_n h_n) - \phi(f)}{t_n} - \phi'(h) \right\|_\bbE = 0
      \end{equation*}
      for all sequences $\{h_n\} \subset \bbD$ and $\{t_n\} \subset \R$ such that $h_n \rightarrow h \in \bbD_0$ and $t_n \rightarrow 0$ as $n \rightarrow \infty$ and $f + t_n h_n \in \bbD_\phi$ for all $n$.
    \item $\phi$ is \emph{Hadamard directionally differentiable} at $f \in \bbD_\phi$ tangentially to a set $\bbD_0 \subseteq \bbD$ if there is a continuous map $\phi'_f: \bbD_0 \rightarrow \bbE$ such that 
      \begin{equation*}
        \lim_{n \rightarrow \infty} \left\| \frac{\phi(f + t_n h_n) - \phi(f)}{t_n} - \phi_{f}'(h) \right\|_\bbE = 0
      \end{equation*}
      for all sequences $\{h_n\} \subset \bbD$ and $\{t_n\} \subset \R_+$ such that $h_n \rightarrow h \in \bbD_0$ and $t_n \searrow 0$ as $n \rightarrow \infty$ and $f + t_n h_n \in \bbD_\phi$ for all $n$.
  \end{enumerate}
\end{deff}
In both cases of the above definition, $\phi'_f$ is continuous, with the addition of linearity in the fully-differentiable case \citep[Proposition 3.1]{Shapiro90}.  They also differ in the sequences of admissible $\{t_n\}$, which allows the second definition to encode directions.

Because the pair of marginal distribution functions always occur as the difference $F_A - F_B$, the next few definitions and lemmas are stated for a single function $f$.  For later results, maps will be applied with the function $f = F_A - F_B$.  The following maps will be used repeatedly in this section and the proofs for analyzing more complex directionally differentiable maps.  Let $\phi: \R \rightarrow \R$ be
\begin{equation}
  \phi(x) = (x)^+ = \max\{ 0, x \},
\end{equation}
and similarly, define $\psi: \R^2 \rightarrow \R$ by
\begin{equation}
  \psi(x, y) = \max\{x, y\}.
\end{equation}
For some domain $\calX \subseteq \R^j$ let $\sigma: \ell^\infty(\calX) \rightarrow \R$ be
\begin{equation} \label{sigma_def}
  \sigma(f) = \sup_{x \in \calX} f(x).
\end{equation}
These are all Hadamard directionally differentiable maps.  It can be verified that for all $a \in \R$,
\begin{equation} \label{phiprime_def}
  \phi'_x(a) = \begin{cases} a & x > 0 \\ \max\{0, a\} & x = 0 \\ 0 & x < 0 \end{cases},
\end{equation}
while for pairs $(a, b) \in \R^2$,
\begin{equation*}
  \psi'_{x,y}(a, b) = \begin{cases} a & x > y \\ \max\{a, b\} & x = y \\ b & x < y \end{cases}.
\end{equation*}
For any $\epsilon \geq 0$, let $\calM_f(\epsilon) = \{x \in \calX : f(x) \geq \sigma(f) - \epsilon\}$ be the set of $\epsilon$-maximizers of $f$.  \citet{CarcamoCuevasRodriguez19} show that for all directions $h \in \ell^\infty(\calX)$
\begin{equation} \label{sigmaprime_def}
  \sigma_f'(h) = \lim_{\epsilon \searrow 0} \sup_{x \in \calM_f(\epsilon)} h(x)
\end{equation}
and they also give conditions under which the limiting operation can be discarded and the supremum of $h$ can be taken over the set of maximizers of $f$.

The next lemma shows shows that a weighted $L_p$ norm (for $p > 1$) applied to the positive part of a function is directionally differentiable.  Cram\'er-von Mises statistics are found by setting $p = 2$.  The directional differentiability of the $L_p$ norm with $p = 1$ was shown in Lemma S.4.5 of \citet{FangSantos19}.  Note that this lemma must be shown for the $L_p$ norm applied to the positive-part map, jointly applied to a function $f$.  This is because $f \mapsto (f)^+$ is not differentiable as a map of functions to functions.  Nevertheless, the dominated convergence theorem allows one to use pointwise convergence with integrability to find the result.

\begin{lemma} \label{lem:lpnorm_hd}
  Suppose $f: \calX \subseteq \R^j \rightarrow \R^k$ is a bounded and $p$-integrable function.  Let $w: \calX \rightarrow \R_+^k$ be such that $\int w_i(x) \dd x < \infty$ for $i = 1, \ldots k$.  Let $1 < p < \infty$ and define the one-sided $L_p$ norm of $f$ by
  \begin{equation} \label{lp_def}
    \lambda(f) = \left( \sum_{i=1}^k \int_\calX \left( (f_i(x))^+ \right)^p w_i(x) \dd x \right)^{1/p}.
  \end{equation}
  For $i = 1, \ldots k$, define the subdomains $\calX_-^i = \{ x \in \calX : f_i(x) < 0 \}$, $\calX_0^i = \{ x \in \calX : f_i(x) = 0 \}$ and $\calX_+^i = \{ x \in \calX : f_i(x) > 0 \}$ and the index collections $\mathcal{I}^0 = \{i \in 1, \ldots k : \mu(\calX_0^i) > 0\}$ and $\mathcal{I}^+ = \{i \in 1, \ldots k : \mu(\calX_+^i) > 0\}$, where $\mu$ is Lebesgue measure.  Then $\lambda$ is Hadamard directionally differentiable and its derivative for any bounded, $p$-integrable $h: \calX \rightarrow \R^k$ is
  \begin{equation} \label{lambdaprime_def}
    \lambda'_f(h) =  \begin{cases}
       0 & \mathcal{I}^+ = \mathcal{I}^0 = \varnothing \\
      \left( \sum_{i \in \mathcal{I}^0} \int_{\calX_0^i} \left( (h_i(x))^+ \right)^p w_i(x) \dd x \right)^{1/p} & \mathcal{I}^+ = \varnothing, \mathcal{I}^0 \neq \varnothing \\
      \frac{1}{\lambda(f)^{p-1}} \sum_{i \in \mathcal{I}^+} \int_{\calX_+^i} f_i^{p-1}(x) h_i(x) w_i(x) \dd x & \mathcal{I}^+ \neq \varnothing
  \end{cases}.
  \end{equation}
  
\end{lemma}

  The above definitions make it easy, if rather abstract, to state the differentiability of the maps from distribution to test statistics that are applied to conduct uniform inference using the $T_1$ process.

\begin{lemma} \label{lem:t1diff}
  Let $f \in \ell^\infty(\calX)$ and let 
  \begin{equation} \label{nu_def}
    \nu(f) = \sup_{x \in \calX} \left( (f(x))^+ + f(-x) \right)^+
  \end{equation}
  and, assuming $f$ is square integrable, 
  \begin{equation} \label{omega_def}
    \omega(f) = \left( \int_\calX \{((f(x))^+ + f(-x))^+\}^2 \dd x \right)^{1/2}.
  \end{equation}
  Then $\nu$ and $\omega$ are Hadamard directionally differentiable, and, letting $f_1(x) = f(-x)$ and $f_2(x) = f(x) + f(-x)$, their derivatives for any direction $h \in \ell^\infty(\calX)$ are
  \begin{equation} \label{nuprime_def}
    \nu_f'(h) = \left( \phi'_{\psi(\sigma(f_1), \sigma(f_2))} \circ \psi'_{\sigma(f_1), \sigma(f_2)} \right) (\sigma_{f_1}'(h), \sigma_{f_2}'(h))
  \end{equation}
  and, assuming in addition that $f, h$ are square integrable,
  \begin{equation} \label{omegaprime_def}
    \omega_f'(h) = \left( \lambda'_{\psi(f_1, f_2)} \circ \psi'_{f_1, f_2} \right) (h, h),
  \end{equation}
  where we take the order $p = 2$ and the weight function $w \equiv 1$ in $\lambda'_f$ defined in~\eqref{lambdaprime_def}.
\end{lemma}

Next we turn to results for the partially identified case.  Lemma~\ref{lem:sup_add} provides the theoretical tool needed for the analysis of Kolmogorov-Smirnov-type statistics when using Makarov bounds.  First define the abstract map $\theta: (\ell^\infty(\calU \times \calX))^2 \rightarrow \R$ by
  \begin{equation} \label{theta_def}
    \theta(f, g) = \sup_{x \in \calX} \left( \sup_{u \in \calU} f(u, x) + \sup_{u \in \calU} g(u, x) \right).
  \end{equation}
  For defining the directional derivative of this map at some $f$ and $g$, we need to consider $\epsilon$-maximizers for any $\epsilon \geq 0$ of these functions in $u$ for each fixed $x$, which for any $f \in \ell^\infty(\calU \times \calX)$ is the set-valued map
  \begin{equation} \label{Af_eps}
    \calM_f(x, \epsilon) = \left\{ u \in \calU : f(u, x) \geq \sup_{u \in \calU} f(u, x) - \epsilon \right\}.
  \end{equation}
  We reserve one special label for the collection of $\epsilon$-maximizers of the outer maximization problem that defines $\theta$: for any $\epsilon \geq 0$ let
  \begin{equation} \label{Btheta_eps}
    \calM_\theta(\epsilon) = \left\{ (u, x) \in \calU \times \calX : f(u, x) + g(u, x) \geq \theta(f, g) - \epsilon \right\}.
  \end{equation}

Lemma~\ref{lem:sup_add} ahead discusses derivatives of $\theta$, a functional that imposes two levels of maximization with an intermediate addition step, and shows that this operator is directionally differentiable.  It is similar to the case of maximizing a bounded bivariate function, and its proof follows that of Theorem 2.1 of \citet{CarcamoCuevasRodriguez19}, which dealt with directional differentiability of the supremum functional applied to a bounded function.  The statement is for the sum of only two functions as arguments but it is straightforward to extend to any finite number of functions, as in Theorem~\ref{thm:partial_teststats}.
\begin{lemma} \label{lem:sup_add}
  Let $\calU \subseteq \R^m$ and $\calX \subseteq \R^n$.  Suppose that $f, g \in \ell^\infty(\calU \times \calX)$, and let $\theta$ be the map defined in~\eqref{theta_def}.  Then $\theta$ is Hadamard directionally differentiable and its derivative at $(f, g)$ for any directions $(h, k) \in (\ell^\infty(\calU \times \calX))^2$ is
  \begin{equation} \label{thetaprime_def}
    \theta'_{f,g}(h, k) = \lim_{\epsilon \searrow 0} \sup_{x \in \calM_\theta(\epsilon)} \left( \sup_{u \in \calM_f(x, \epsilon)} h(u, x) + \sup_{u \in \calM_g(x, \epsilon)} k(u, x) \right).
  \end{equation}
\end{lemma}

\subsection{Proof of results in Appendix~\ref{app:hadamard}} \label{app:appA_proof}

\begin{proof}[Proof of Lemma~\ref{lem:lpnorm_hd}]
  Let $\{t_n\}$ be a sequence of positive numbers such that $t_n \searrow 0$ as $n \rightarrow \infty$, and let $\{h_n\} \in (\ell^\infty(\calX))^k$ be a sequence of bounded, $p$-integrable functions such that $h_n \rightarrow h \in (\ell^\infty(\calX))^k$ as $n \rightarrow \infty$.  

  Suppose that for all $i$ and all $x \in \calX$, $f_i(x) < 0$, or in other words, $\mathcal{I}^+ = \mathcal{I}^0 = \varnothing$.  For any point $x$ there exists some $N$ such that for all $n > N$, $(f_i + t_n h_{ni})^+ = 0$ because $t_n \searrow 0$ and $h_i$ is bounded.  Then dominated convergence implies that the $p$-th power of the $L_p$ norm satisfies
  \begin{equation*}
    \lim_{n \rightarrow \infty} \frac{1}{t_n} \Bigg( \sum_{i=1}^k \int_{\calX_-^i} \left( (f_i(x) + t_n h_{ni}(x))^+ \right)^p w_i(x) \dd x - \sum_{i=1}^k \int_{\calX_-^i} \left( (f_i(x))^+ \right)^p w_i(x) \dd x \Bigg) = 0.
  \end{equation*}
  This is also the result for $\lambda(f)$ in this case, which is the difference of these terms each raised to the power $1/p$.

  Next suppose $\mathcal{I}^0 \neq \varnothing$ and $\mathcal{I}^+ = \varnothing$, that is, for some $i$, $\{\calX_0^i\}$ has positive measure but the measure of $x$ that make any coordinate of $f$ positive is zero.  Then calculate the differences directly:
  \begin{align*}
    \lim_{n \rightarrow \infty} \frac{1}{t_n} \Bigg\{ \Bigg( \sum_{i=1}^k \int_{\calX_0^i} \big( (f_i(x) &+ t_n h_{ni}(x))^+ \big)^p w_i(x) \dd x \Bigg)^{1/p} - \left( \sum_{i=1}^k \int_{\calX_0^i} \left( (f_i(x))^+ \right)^p w_i(x) \dd x \right)^{1/p} \Bigg\} \\
    {} &= \lim_{n \rightarrow \infty} \frac{1}{t_n} \left( t_n^p \sum_{i=1}^k \int_{\calX_0^i} \left( (h_{ni}(x))^+ \right)^p w_i(x) \dd x \right)^{1/p} \\
    {} &= \sum_{i=1}^k \int_{\calX_0^i} \left( (h_{i}(x))^+ \right)^p w_i(x) \dd x 
  \end{align*}
  using dominated convergence and the $p$-integrability of $h$.  If the subregions $\{x : f_i(x) < 0\}$ have positive measure, they contribute 0 to the limit.

  Now suppose that $\mathcal{I}^+$ is not empty, that is, there is at least one $i$ such that $\calX_+^i$ has positive measure.  Then for each $x \in \calX_i^+$ there exists an $N$ such that for $n > N$, $f_i(x) + t_n h_{ni}(x) > 0$ for all $i$.  Then for $n > N$, for this $x$,
  \begin{align*}
    (f_i(x) + t_n h_{ni}(x))^p - f_i^p(x) &= \sum_{j=0}^p \binom{p}{j} f_i^j(x) (t_n h_{ni}(x))^{p-j} - f_i^p(x) \\
    {} &= f_i^p(x) + p t_n f_i^{p-1}(x) h_{ni}(x) + O(t_n^2) - f_i^p(x) \\
    {} &= p t_n f_i^{p-1}(x) h_{ni}(x) + O(t_n^2).
  \end{align*}
  This implies that for $n$ large enough, the inner integral, using the calculations from the previous parts to account for the sets where $f_i$ is zero or negative, satisfies
  \begin{align*}
    \lim_{n \rightarrow \infty} &\frac{1}{t_n} \left\{ \sum_{i=1}^k \int_{\calX} (f_i(x) + t_n h_{ni}(x))^p w_i(x) \dd x - \sum_{i=1}^k \int_{\calX} f_i^p(x) w_i(x) \dd x \right\} \\
    &= \lim_{n \rightarrow \infty} \frac{1}{t_n} \left\{ p t_n \sum_{i=1}^k \int_{\calX_+^i} f_i^{p-1}(x) h_{ni}(x) w_i(x) \dd x + O(t_n^2) + O(t_n^p) + 0 \right\} \\
    &= p \sum_{i=1}^k \int_{\calX_+^i} f_i^{p-1}(x) h_{i}(x) w_i(x) \dd x.
  \end{align*}
    Using the expansion $(x + th_t)^{1/p} = x^{1/p} + \frac{1}{p} x^{(1-p)/p} th_t + o(|th_t|)$ as $t \rightarrow 0$, it can be seen that the Hadamard derivative of $x \mapsto x^{1/p}$ is $\frac{1}{p} x^{(1-p)/p} h$.  Therefore the chain rule and integrability of $f$ and $h$ implies that the derivative is
  \begin{equation*}
    \frac{1}{\lambda(f)^{p-1}} \sum_{i=1}^k \int_{\calX_+^i} f_i^{p-1}(x) h_{i}(x) w_i(x) \dd x.
  \end{equation*}
\end{proof}

\begin{proof}[Proof of Lemma~\ref{lem:t1diff}]
  For $\nu$ write
  \begin{align*}
   \nu(f) &= \sup_{x \in \calX} \left( (f(x))^+ + f(-x) \right)^+ \\
   {} &= \sup_{x \in \calX} \max \left\{ 0, (f(x))^+ + f(-x) \right\} \\
    {} &= \sup_{x \in \calX} \max \left\{ 0, \max \left\{ f(-x), f(x) + f(-x) \right\} \right\} \\
    \intertext{and using the definitions of $f_1$ and $f_2$ made in the statement of the lemma and changing the order in which the maxima are computed}
    {} &= \max \left\{ 0, \max \left\{ \sup_{x \in \calX} f_1(x), \sup_{x \in \calX} f_2(x) \right\} \right\} \\
    {} &= (\phi \circ \psi) (\sigma(f_1), \sigma(f_2)).
  \end{align*}
  Then using the chain rule \citep{Shapiro90} the derivative is that given in the statement of the lemma.  For $\omega$, assume $f$ and $h$ are square integrable and write
  \begin{align*}
    \omega(f) &= \lambda((f(x))^+ + f(-x)) \\
    {} &= \lambda( \max\{ f(-x), f(x) + f(-x) \} ) \\
    {} &= (\lambda \circ \psi) (f_1, f_2).
  \end{align*}
  Taking a derivative and using the chain rule implies the second expression in the statement of the lemma.
\end{proof}

\begin{proof}[Proof of Lemma~\ref{lem:sup_add}]
  First, let $s_n = t_n^{-1}$ and define the finite differences  
  \begin{equation}
    \Delta_n = \sup_\calX \left( \sup_\calU (s_n f + h)(u, x) + \sup_\calU (s_n g + k)(u, x) \right) - s_n \theta(f, g)
  \end{equation}
  so that for any $s_n \nearrow \infty$, we need to show that $\Delta_n \rightarrow \theta'_{f, g}(h, k)$ defined in the statement of the theorem.

  Fix an $\epsilon > 0$.  Then for any $x \notin \calM_\theta(\epsilon)$, note that
  \begin{equation}
    \sup_\calU (s_n f + h)(u, x) + \sup_\calU (s_n g + k)(u, x) - s_n \theta(f, g) \leq \sup h + \sup k - s_n \epsilon.
  \end{equation}
  Similarly, if $u \notin \calM_f(x, \epsilon)$ for any $x$ (the case for $u$ that do not nearly-optimize $g(\cdot, x)$ is symmetric), then also
  \begin{equation}
    (s_n f + h)(u, x) + \sup_\calU (s_n g + k)(u, x) - s_n \theta(f, g) \leq \sup h + \sup k - s_n \epsilon
  \end{equation}
  for that $x$.  Therefore for any $\epsilon > 0$,
  \begin{multline}
    \limsup_n \Delta_n \\
     = \limsup_n \left( \sup_{\calM_\theta(\epsilon)} \left( \sup_{\calM_f(x, \epsilon)} (s_n f + h)(u, x) + \sup_{\calM_g(x, \epsilon)} (s_n g + k)(u, x) \right) - s_n \theta(f, g) \right) \\
    {} \leq \limsup_n \Bigg( s_n \sup_{\calM_\theta(\epsilon)} \left( \sup_{\calM_f(x, \epsilon)} f(u, x) + \sup_{\calM_g(x, \epsilon)} g(u, x) \right) - s_n \theta(f, g) \\
    + \sup_{\calM_\theta(\epsilon)} \left( \sup_{\calM_f(x, \epsilon)} h(u, x) + \sup_{\calM_g(x, \epsilon)} k(u, x) \right) \Bigg) \\
    {} = \sup_{\calM_\theta(\epsilon)} \left( \sup_{\calM_f(x, \epsilon)} h(u, x) + \sup_{\calM_g(x, \epsilon)} k(u, x) \right),
  \end{multline}
  so that this inequality holds as $\epsilon \searrow 0$.

  Next, for any $\epsilon > 0$ define
  \begin{equation}
    \bar{t}(\epsilon) = \sup_{\calM_\theta(\epsilon)} \left( \sup_{\calM_f(x, \epsilon)} h(u, x) + \sup_{\calM_g(x, \epsilon)} k(u, x) \right).
  \end{equation}
  Because this function is nondecreasing in $\epsilon$, it has a limit as $\epsilon \searrow 0$, so that for any $m \in \mathbb{N}$ there exists an $x_m \in \calM_\theta(1/m)$ and $(u_m^f, u_m^g)$ satisfying the inequality
  \begin{equation*}
    h(u_m^f, x_m) + k(u_m^g, x_m) \geq \bar{t}(1/m) - 1/m.
  \end{equation*}
  Therefore
  \begin{multline}
    \bar{t}(1/m) \leq h(u_m^f, x_m) + k(u_m^g, x_m) + 1/m \\
    = s_n f(u_m^f, x_m) + h(u_m^f, x_m) + s_n g(u_m^g, x_m) + k(u_m^g, x_m) \\
    + 1/m - s_n(f(u_m^f, x_m) + g(u_m^g, x_m)) \\
    \leq \sup_\calX \left( \sup_\calU (s_n f + h)(u, x) + \sup_\calU (s_n g + k)(u, x) \right) - s_n \theta(f, g) + (s_n + 1) / m,
  \end{multline}
  which implies that
  \begin{equation}
    \lim_{\epsilon \searrow 0} \sup_{\calM_\theta(\epsilon)} \left( \sup_{\calM_f(x, \epsilon)} h(u, x) + \sup_{\calM_g(x, \epsilon)} k(u, x) \right) = \lim_{m \rightarrow \infty} \bar{t}(1/m) \leq \Delta_n.
  \end{equation}
\end{proof}

\begin{proof}[Proof of Theorem~\ref{thm:resample_size}]
  This is an application of Corollary 3.2 in \citet{FangSantos19}, and we only sketch the most important details of the proof.  After applying the null hypothesis, the derivatives $\nu_F'$ and $\omega_F'$ shown in~\eqref{nuprime_def} and~\eqref{omegaprime_def} are both convex.  For example, in the expression for $\nu_F'$,
  \begin{equation*}
    \left( \sup_{\calX_0^1(P)} (\alpha h_{1A} + (1 - \alpha) h_{1B}) \right)^+ \leq \alpha \left( \sup_{\calX_0^1(P)} h_{1A} \right)^+ + (1 - \alpha) \left( \sup_{\calX_0^1(P)} h_{1B} \right)^+ 
  \end{equation*}
  and similar calculations hold for the other two terms.  In the case of $\omega_F'$, for example,
  \begin{equation*}
    \int_{\calX_0^1(P)} \left( \left( \alpha h_1 + (1 - \alpha) h_2 \right)^+ \right)^2 \leq \alpha \int_{\calX_0^1(P)} \left( \left( h_1 \right)^+ \right)^2 + (1 - \alpha) \int_{\calX_0^1(P)} \left( \left( h_2 \right)^+ \right)^2,
  \end{equation*}
  where the inequality relies on the nonnegativity of the innermost term and convexity of $x \mapsto x^2$ for $x \geq 0$.  Then Theorem 3.3 of \citet{FangSantos19} applies.  The second part of the theorem is a special case of the first, when the part of the relationship that leads to nondegenerate behavior is not empty.
\end{proof}

\begin{proof}[Proof of Theorem~\ref{thm:resample_size_intervalID}]
  Start by considering $V_3$.  As in the proof of Theorem~\ref{thm:resample_consistent}, we simplify the analysis by writing this statistic as a composition of maps that act on just two functional arguments, $(\phi'_{\theta(f,g)} \circ \theta'_{f,g}) (h,k)$, where the positive-part map $\phi'_x$ is defined in~\eqref{phiprime_def} and $\theta'_{f,g}$ is, for any $h,k \in \ell^\infty(\calU \times \calX))$, 
  \begin{equation*}
    \theta'_{f,g}(h, k) = \lim_{\epsilon \searrow 0} \sup_{x \in \calM_\theta(\epsilon)} \left( \lim_{\epsilon \searrow 0} \sup_{u \in \calM_f(x, \epsilon)} h(u, x) + \lim_{\epsilon \searrow 0} \sup_{u \in \calM_g(x, \epsilon)} k(u, x) \right),
  \end{equation*}
  where $\calM_f(x, \epsilon)$ and $\calM_\theta(\epsilon)$ are defined in~\eqref{Af_eps} and~\eqref{Btheta_eps}.  
  
  It can be verified that for a fixed value of $\theta(f,g)$, $\hat{\phi}'_{\theta(f, g)}(x)$ is convex and nondecreasing.  Next consider $\theta'_{f,g}$.  For any $\epsilon > 0$, consider the map applied to the convex combination of vector-valued functions $\alpha (h_1, k_1) + (1 - \alpha) (h_2, k_2)$:
  \begin{multline*}
    \sup_{\calM_\theta(\epsilon)} \left( \sup_{\calM_f(x, \epsilon)} (\alpha h_1(u, x) + (1 - \alpha) k_1(u, x)) + \sup_{\calM_g(x, \epsilon)} (\alpha h_2(u, x) + (1 - \alpha) k_2(u, x)) \right) \\
    \leq \sup_{\calM_\theta(\epsilon)} \left( \alpha \left( \sup_{\calM_f(x, \epsilon)} h_1(u, x) + \sup_{\calM_g(x, \epsilon)} k_1(u, x) \right) + (1 - \alpha) \left( \sup_{\calM_f(x, \epsilon)} h_2(u, x) + \sup_{\calM_g(x, \epsilon)} k_2(u, x) \right) \right) \\
    \leq \alpha \sup_{\calM_\theta(\epsilon)} \left( \sup_{\calM_f(x, \epsilon)} h_1(u, x) + \sup_{\calM_g(x, \epsilon)} k_1(u, x) \right) + (1 - \alpha) \sup_{\calM_\theta(\epsilon)} \left( \sup_{\calM_f(x, \epsilon)} h_2(u, x) + \sup_{\calM_g(x, \epsilon)} k_2(u, x) \right).
  \end{multline*}
  Therefore, letting $\epsilon \searrow 0$, it can be seen that $\theta'_{f,g}$ is convex.  Because $V_3$ is the composition of a non-decreasing, convex function with a convex function, $V_3$ is also a convex map of $(h, k)$ to $\R$ \citep[eq. 3.11]{BoydVandenberghe04}.  As mentioned in the text, $\calP_0 \subseteq \calP^{nec}$.  Therefore Corollary 3.2 of \citet{FangSantos19} implies 
\begin{equation*}
  \limsup_{n \rightarrow \infty} P_n \left\{ V_{3n} > q_{V_3^*}(1 - \alpha) \right\} \leq \alpha.
\end{equation*}

  Turn next to $W_3$.  Similarly, write this statistic as a map of pairs of bounded functions to the real line as $W_{3n} = (\lambda'_{\mu(f, g)} \circ \mu'_{f,g})(h, k)$, where for each $x \in \calX$,
  \begin{equation*}
    \mu(f, g)(x) = \sup_\calU f(u, x) + \sup_\calU g(u, x)
  \end{equation*}
  and 
  \begin{equation*}
    \mu_{f,g}'(h,k)(x) = \lim_{\epsilon \searrow 0} \max_{u \in \calM_f(x, \epsilon)} h(u, x) + \lim_{\epsilon \searrow 0} \max_{u \in \calM_g(x, \epsilon)} k(u, x),
  \end{equation*}
  and for any functions $f, h \in \ell^\infty(\calX)$, $\lambda'_f(h)$ is defined in~\eqref{lambdaprime_def}.  We show the convexity of this composition directly.  Paraphrase $\mu(x) = \mu(f, g)(x)$, and for fixed $\epsilon > 0$, 
  \begin{align*}
    \mu_1'(x) &= \sup_{u \in \calM_f(x, \epsilon)} h_1(u, x) + \sup_{u \in \calM_g(x, \epsilon)} k_1(u, x) \\
    \mu_2'(x) &= \sup_{u \in \calM_f(x, \epsilon)} h_2(u, x) + \sup_{u \in \calM_g(x, \epsilon)} k_2(u, x) \\
    \bar{\mu}'(x) &= \sup_{u \in \calM_f(x, \epsilon)} (\alpha h_1 + (1 - \alpha) k_1)(u, x) + \sup_{u \in \calM_g(x, \epsilon)} (\alpha h_2 + (1 - \alpha) k_2)(u, x).
  \end{align*}
    Finally, let $\calX_0$ denote the region where $\mu(x) = 0$.  Then Lemma~\ref{lem:lpnorm_hd} shows that $\lambda_\mu'(\bar{\mu}') = \lambda(\bar{\mu}'|_{\calX_0})$, where $\bar{\mu}'|_{\calX_0}$ denotes the restriction of the function $\bar{\mu}'$ to the set $\calX_0$.  Consider the first term on the right hand side.  Inside the integral, it can be seen that
  \begin{align*}
    0 &\leq \left( \bar{\mu}'(x) \right)^+ \\
    {} &= \left( \sup_{u \in \calM_f(x, \epsilon)} (\alpha h_1 + (1 - \alpha) h_2)(u, x) + \sup_{u \in \calM_g(x, \epsilon)} (\alpha k_1 + (1 - \alpha) k_2)(u, x) \right)^+ \\
    {} &\leq \Bigg( \alpha \left( \sup_{u \in \calM_f(x, \epsilon)} h_1(u, x) + \sup_{u \in \calM_g(x, \epsilon)} k_1(u, x) \right) \\
    {} &\phantom{=} \qquad \qquad + (1 - \alpha) \left( \sup_{u \in \calM_f(x, \epsilon)} h_2(u, x) + \sup_{u \in \calM_g(x, \epsilon)} k_2(u, x) \right) \Bigg)^+ \\
    {} &= \left( \alpha \mu_1'(x) + (1 - \alpha) \mu_2'(x) \right)^+ \\
    {} &\leq \alpha \left( \mu_1'(x) \right)^+ + (1 - \alpha) \left( \mu_2'(x) \right)^+.
  \end{align*}
  Because the integrand is nonnegative, subadditivity of the $L_2$ norm implies 
  \begin{equation*}
    \lambda(\bar{\mu}'|_{\calX_0}) \leq \alpha \lambda(\mu_1'|_{\calX_0}) + (1 - \alpha) \lambda(\mu_2'|_{\calX_0}).
  \end{equation*}
  This inequality holds as $\epsilon \searrow 0$ by the assumed square-integrability of the arguments.  Therefore Corollary 3.2 of \citet{FangSantos19} implies 
  \begin{equation*}
    \limsup_{n \rightarrow \infty} P_n \left\{ W_{3n} > q_{W_3^*}(1 - \alpha) \right\} \leq \alpha.
  \end{equation*}
\end{proof}

%% file: supp_extras/all_tests.tex
%latex.default(tab, file = "all_tests.tex", label = "tab:all_pointIDtests",     rowname = testpv, title = "", ctable = TRUE, math.col.names = TRUE,     col.just = rep("c", 2), cgroup = c("LASD", "FOSD"), n.cgroup = c(1,         1), colheads = hypname, rgroup = timename, n.rgroup = c(4,         4), caption.loc = "bottom", caption = "Supremum and $L_2$ tests for inferring whether the Jobs First \n      (JF) program would be preferred to the Aid to Families with Dependent \n      Children (AFDC).  The first column uses changes in income and the second \n      column measures income in levels without regard to pre-policy income.  \n      Both types of test statistic agree on rejection decisions.  1999 \n      bootstrap repetitions used in each test.")%
\ctable[botcap,caption={Supremum and $L_2$ tests for inferring whether the Jobs First 
      (JF) program would be preferred to the Aid to Families with Dependent 
      Children (AFDC).  The first column uses changes in income and the second 
      column measures income in levels without regard to pre-policy income.  
      Both types of test statistic agree on rejection decisions.  1999 
      bootstrap repetitions used in each test.},label=tab:all_pointIDtests,pos=!tbp,]{lccc}{}{\FL
\multicolumn{1}{l}{\bfseries }&\multicolumn{1}{c}{\bfseries LASD}&\multicolumn{1}{c}{\bfseries }&\multicolumn{1}{c}{\bfseries FOSD}\NN
\cline{2-2} \cline{4-4}
\multicolumn{1}{l}{}&\multicolumn{1}{c}{$F_{JF} \succeq F_{AFDC}$}&\multicolumn{1}{c}{}&\multicolumn{1}{c}{$G_{JF} \succeq G_{AFDC}$}\ML
{\bfseries Before JF time limit}&&&\NN
~~supremum norm&$ 0.3283$&&$0.7006$\NN
~~p-value&$ 0.8954$&&$0.7484$\NN
~~L2 norm&$ 0.1790$&&$0.2240$\NN
~~p-value&$ 0.9095$&&$0.8634$\ML
{\bfseries After JF time limit}&&&\NN
~~supremum norm&$12.9315$&&$1.5446$\NN
~~p-value&$ 0.0000$&&$0.2801$\NN
~~L2 norm&$ 9.1380$&&$2.1285$\NN
~~p-value&$ 0.0000$&&$0.1436$\LL
}